\documentclass[11pt]{article}
\usepackage[margin=1in]{geometry}
\usepackage[utf8]{inputenc}
\usepackage[hidelinks]{hyperref}
\usepackage{xcolor}
\usepackage{amsthm}
\usepackage{amssymb, amsmath}
\usepackage[sort&compress, numbers]{natbib}
\usepackage{bm}
\usepackage{tikz}
\usepackage{tikz-3dplot}
\usetikzlibrary{calc,math}
\usepackage{booktabs}
\usepackage{url}
\usepackage{setspace}
\usepackage[export]{adjustbox}
\usepackage{placeins}
\usepackage{authblk}

\definecolor{lightred}{RGB}{251,154,153}
\definecolor{darkblue}{RGB}{31,120,180}
\definecolor{lightgreen}{RGB}{178,223,138}
\definecolor{lightpink}{RGB}{251,154,153}
\definecolor{darkred}{RGB}{227,26,28}
\definecolor{darkgreen}{RGB}{51,160,44}
\definecolor{lightblue}{RGB}{166,206,227}
\definecolor{lila}{RGB}{106,61,154}

\usepackage{pgfplots}
\pgfplotsset{compat=1.18}
\usepackage{pgfplotstable}

\newcommand{\ppsi}{\bm{\psi}}

\DeclareMathOperator{\End}{End}

\newcommand{\mcA}{\mathcal{A}}

\newcommand{\mcN}{\mathcal{N}}
\newcommand{\mcO}{\mathcal{O}}

\newcommand{\ket}[1]{\left|#1\right\rangle}
\newcommand{\bra}[1]{\left\langle#1\right|}
\newcommand{\Tr}[1]{\operatorname{Tr}\!\left(#1\right)}

\title{3d Conformal Field Theories via Fuzzy Sphere Algebra}

\author[1]{Luisa Eck}
\author[2]{Zhenghan Wang}

\affil[1]{\small Walter Burke Institute for Theoretical Physics and Department of Physics,
California Institute of Technology, Pasadena, CA, 91125, USA\\
\texttt{luisa@caltech.edu}}
\affil[2]{\small Department of Mathematics, University of California, Santa Barbara, CA 93106, USA\\
\texttt{zhenghwa@math.ucsb.edu}}

\begin{document}

\maketitle

\begin{abstract}
    Fuzzy sphere models conjecturally realize 3d CFTs in small systems of spinful fermions, but why they work so well is still not fully understood. Their Hamiltonians are built from electron density operators projected to the lowest Landau level. 
    We analyze the Lie algebra generated by these density modes and its large-$s$ limits. Depending on how the limit is taken, the algebra approaches either the Girvin–MacDonald–Platzman algebra in a local planar limit or a semiclassical algebra for low-angular-momentum modes in a global commutative limit. With an additional restriction to a low-excitation sector above the paramagnetic state, the density modes become approximate harmonic oscillators. We also test whether the conformal algebra $so(3,2)$ can be realized directly by density modes. Such a representation exists only in the minimal two-electron system; its natural coproduct extension does not match the physical thermodynamic limit of a single growing fuzzy sphere.
\end{abstract}

\section{Introduction}

3d topological quantum field theories (TQFTs) and 2d conformal field theories (CFTs) are among the best examples of quantum field theories that are both mathematically rigorous and physically relevant.  They are related by an edge-bulk correspondence that is realized by quantum Hall materials. 
A generalization of the 3d TQFT---2d CFT correspondence to one dimension higher would be both interesting and daunting. 4d TQFTs are expected to describe topological insulators and superconductors, while 2+1d critical points, such as the phase transition in the transverse field Ising model, are believed to be governed by 3d CFTs.

One possible mathematical approach to 2d CFTs, in particular the minimal models, is via anyonic chains \cite{zini2018conformal}.  This generalized spin chain approach is based on finite-dimensional Hilbert space approximations, hence obviously mathematical, but subtleties arise in the limiting process.  Extensive numerical simulation and mathematical verification of simple examples provide strong evidence that this anyonic chain approach to 2d CFTs, at least the minimal models, is viable \cite{feiguin2007interacting, zini2018conformal}. 

There are two important classes of algebras in the anyonic chain approach to 2d CFTs: the finite-dimensional Temperley-Lieb-Jones (TLJ) algebras $TLJ_n(A)$, and the infinite-dimensional Virasoro algebras.  Both algebras play essential roles in the solution and understanding of the 2d Ising model and CFTs.  At certain special values of the Kauffman variable $A$, the quotients of the Temperley-Lieb (TL) algebras $TL_n(A)$ by the Jones-Wenzl projectors lead to $\mathbb{C}^*$-TLJ algebras $TLJ_n(A)$ generated by TL generators $\{e_i\}_{i=1}^{n-1}$ \cite{wang2010topological}.  These TL generators $\{e_i\}$ could be regarded as finite lattice version of the Virasoro generators $\{L_m\}$ in the sense given by the Koo-Saleur formulas via the Fourier transform \cite{koo1994representations,zini2018conformal}. 

A natural question is what should play the role of anyonic chains in 2+1 dimensions for the purpose of realizing 3d CFTs. 
Anyonic chains are built from fusion categories, and one possible higher-dimensional generalization is provided by fusion surface models \cite{inamura2024fusion,Eck:2024myo}, which are built from fusion 2-categories. However, these models appear more naturally suited for topological phases than for critical points. In this work, we instead regard fuzzy sphere Hamiltonians as 2+1d analogues of generalized spin chains.

Fuzzy sphere models, introduced in \cite{he2022fuzzy}, are built from spinful fermions in the lowest Landau level on a sphere. Numerical studies have found that their low-energy spectra and correlation functions can approximate 3d CFT data with surprisingly small system sizes.  
We emphasize, however, that obtaining CFT-like spectra from small 2+1d systems is not unique to the fuzzy sphere: similar behavior occurs in more conventional discretizations, such as the 2+1d transverse-field Ising model on the icosahedron \cite{Lao:2023zis}, where good accuracy requires conformal perturbation theory. Conformal perturbation theory is likewise useful on the fuzzy sphere to further improve the extracted scaling dimensions \cite{Lauchli:2025fii, voinea2025anyons}.  

The main object of this paper is the algebra generated by the Landau-level-projected density modes that enter fuzzy sphere Hamiltonians. We first review the fuzzy sphere construction in Sec.~\ref{sec:fuzzyreview}, including the Hamiltonians, their symmetries, and the density modes used in previous work. The density mode commutator itself is also known, and we recall it in Sec.~\ref{subsec:commrel}. Our new contribution begins by analyzing this commutator as an algebraic object in its own right: we relate the density modes to fuzzy spherical harmonics from noncommutative field theory \cite{Madore:1991bw,Grosse:1995ar} and to the matrix Lie algebra $su(2N)$, prove that the abstract bracket satisfies the Jacobi identity, and identify commuting subalgebras in Sec.~\ref{subsec:jacobicommuting}.

We then ask what this density algebra becomes in the large-$s$ regimes relevant for continuum limits. In Sec.~\ref{sec:thermodymlimit}, we distinguish two natural limits: a local planar limit, where modes with large angular momentum recover the fuzzy plane/GMP algebra, and a small-angular-momentum semiclassical limit. Within this semiclassical regime, we also make precise the oscillator approximation observed in Ref.~\cite{He2025scalar}: after restricting to a low-excitation sector above the paramagnetic state, the density modes behave approximately as independent harmonic oscillators.

Finally, we ask whether the conformal algebra $so(3,2)$ itself, reviewed in Sec.~\ref{sec:cft}, can be realized directly by the density modes.
If possible, this would suggest that representation-theoretic methods may be useful for approaching 3d CFTs. In Sec.~\ref{sec: Hopf}, we find such a representation only in the minimal two-electron fuzzy sphere. The natural \(so(3)\)-equivariant coproduct extends it to larger tensor-product Hilbert spaces, but not to the physical thermodynamic limit of a single growing fuzzy sphere.

Our results clarify algebraic structures that may enter a future low-energy scaling limit of fuzzy sphere models realizing 3d CFTs. A controlled scaling limit is needed to turn the observed finite-size agreement with CFT spectra and correlation functions into a genuine convergence statement. Since such a scaling limit is difficult and subtle to take, we leave it to the future. 
A related long-term goal is to identify higher-dimensional analogues of the TLJ and Virasoro algebras that organize \(1+1\)d lattice realizations of CFTs. Such structures, if they exist, could provide new guidance for locating critical points.
We speculate that these symmetry algebras are potentially connected to diffeomorphisms of $S^2 \times S^1$ -- the 2-fold cover of the conformal compactification of $\mathbb{R}^{2,1}$.

\section{3d CFTs}\label{sec:cft}

Above two dimensions, the conformal transformations are always global and only form finite-dimensional Lie groups. For Euclidean spacetime of dimension $n$, the conformal group is $SO(n+1,1)$, while for the Lorentzian signature $(n-1,1)$, the conformal group is $SO(n,2)$.  We are interested in $n=3$ and Minkowski spacetime (as the fuzzy sphere models are formulated as 2+1d quantum Hamiltonians), for which the relevant conformal algebra is $so(3,2)$.

\subsection{Conformal invariance and covariance}

A 3d (quantum) conformal field theory (CFT) is a quantum field theory with symmetry extending the Poincare group to one that includes scale and M\"obius transformations.  While the correlation functions of a CFT are invariant, the fields and states only transform covariantly under the action of the extended symmetry group.  Therefore, representation theory of the extended symmetry group could be a powerful tool for constructing mathematically nontrivial CFTs, especially when the symmetry is infinite-dimensional, as in two dimensions.

\subsection{Conformal \texorpdfstring{$so(3,2)$}{so(3,2)} algebra}

The $so(3,2)$ algebra has 10 antisymmetric generators $M_{AB} = - M_{BA}$ with $A,B \in \{1,2,3,4,5\}$. Two generators commute if all indices are distinct, $[M_{AB}, M_{CD}]=0$. The nontrivial commutation relations are 
\begin{align*}
    [M_{AB}, M_{AC}] = \begin{cases}
        i M_{BC} & \text{ if } A = 4,5 \\ 
        -i M_{BC} & \text{ if } A = 1,2,3 \\ 
    \end{cases}.
\end{align*}
In particular, the $so(3)$ rotation subalgebra is generated by
\begin{align*}
    J_z = M_{12}, \quad J_\pm = M_{23} \pm i M_{13}.
\end{align*}
A standard physics basis for the remaining generators is given by the dilatation $D$, translations $P_A$ and special conformal transformations $K_A$ ($A=1,2,3$):
\begin{align}\label{eq:confgenerators}
    D = M_{45}, \quad P_A = M_{4A} + i M_{5A}, \quad K_A = M_{4A} - i M_{5A}.
\end{align}

Following \cite{celeghini2013algebraic}, it is sometimes convenient to instead use two commuting Cartans and four pairs of ladder operators as generators of $so(3,2)$. We denote 
\begin{align*}
    J_z &= M_{12}, \quad J_\pm = M_{23} \pm i M_{13}, \quad D_z = M_{45}, \quad D_\pm = M_{43} \pm i M_{53},
\end{align*}
where $D_z$ coincides with the dilatation operator $D$ and $D_\pm$ correspond to $K_z$ and $P_z$ in the usual physics convention \eqref{eq:confgenerators}). The remaining four ladder operators are defined by commutators,
\begin{align*}
    R_\pm = \pm [J_\pm, D_\pm], \quad S_\pm = \pm [J_\mp, D_\pm].
\end{align*}
In this basis, the commutation relations are given by
\begin{equation}\label{eq:so32commrel}
    \begin{aligned}
        [J_z, J_{\pm}] &= \pm J_\pm, \quad [J_+, J_-] = 2J_z, \quad [D_z, D_{\pm}] = \pm D_\pm, \quad [D_+, D_-] = -2D_z, \\
        [D_z, Jz] &= 0, \quad [D_z, J_\pm] = 0, \quad [D_\pm, J_z] = 0 \\ 
        [D_\mp, R_\pm] &= \pm 2J_\pm, \quad [J_\pm, S_\pm] = \pm 2D_\pm \quad \text{ for } R_\pm = \pm [J_\pm, D_\pm], \; S_\pm = \pm [J_\mp, D_\pm], \\
        [D_\pm, R_\pm] &= 0, \quad [J_\pm, R_\pm] = 0, \quad [D_\pm, S_\pm] = 0, \quad [J_\mp, S_\pm] = 0.
    \end{aligned}
\end{equation}
All remaining commutators follow from the ones above and the Jacobi identity.

\subsection{Representations of \texorpdfstring{$so(3,2)$}{so(3,2)}}

Because $so(3,2)$ is a noncompact Lie algebra, its finite-dimensional representations are all non-unitary. Unitary CFTs instead use infinite-dimensional, positive-energy unitary irreducible representations; an example of such is given by Dirac in \cite{dirac1963remarkable}. 

For a critical quantum Hamiltonian on the spatial sphere, emergent conformal symmetry is visible in the low-energy spectrum. In radial quantization, time translations are generated by the dilatation operator $D$, so scaling dimensions $\Delta$ are the eigenvalues of $D$. Via the state–operator correspondence, low-lying energy eigenstates correspond to local CFT operators labeled by $(\Delta, l)$: $l$ is the $so(3)$ angular momentum, and the energy is proportional to $\Delta$.

A useful way to understand this unitarity obstruction is to recall that the real Lie algebras $so(p,q)$ and $so(p+q)$ become isomorphic as complex Lie algebras after complexification, $so(p,q)_\mathbb{C} \cong so(p+q)_\mathbb{C}$. As a result, the same finite-dimensional complex representation can be viewed either as a representation of the non-compact real form $so(p,q)$ or the compact real form $so(p+q)$. While it is necessarily non-unitary for $so(p,q)$, it can admit a positive-definite invariant inner product—hence become unitary—when interpreted as a representation of $so(p+q)$. We illustrate this point below using the pair $so(2,1)$ and $so(3)$. 

Both Lie algebras are generated by $\{L_i\}_{i=1}^3$ with Lie brackets
$$
    [L_1,L_2] =i\sigma L_3, \quad [L_2,L_3] = iL_1, \quad [L_3,L_1] = iL_2,
$$
with $\sigma=+1$ for $so(3)$ and $\sigma=-1$ for $so(2,1)$.  Set $H = L_3, E_{\pm}= L_1 \pm iL_2$, then the Casimir is $$C = \sigma E_{\mp}E_{\pm} + H(H \pm 1).$$ The Casimir $C$ is a constant $c$ for an irreducible representation.  In a unitary representation, $E_- = (E_+)^\dagger$ and $\bra{m} E_- E_+ \ket{m} = || E_+ \ket{m}||^2 \geq 0$ for any $H$-eigenstate $H \ket{m} = m \ket{m}$. Applying the Casimir relation to $\ket{m}$ yields
$$
\sigma || E_+ \ket{m}||^2 = c - m(m+1).
$$
For $\sigma=+1$, the right hand side eventually becomes negative as $m$ gets too large in magnitude, which forces the ladder of $H$ eigenvalues to terminate at both ends and gives a finite-dimensional unitary representation (with $m=-j,\dots,j$ at fixed $c=j(j+1)$). 
In contrast, for $\sigma=-1$, the right hand side grows as $m^2$, so there is no analogous obstruction: the ladder can extend indefinitely, and any nontrivial unitary representation must be infinite-dimensional.

\subsection{CFTs as scaling limits of generalized spin chains}\label{subsec:scalinglimit}

Anyonic chains are generalized spin chains in 1+1 dimensions.  A key motivation for studying them is their convergence to minimal model CFTs in their scaling limits \cite{zini2018conformal}.  While numerical simulations provide physical proof of their convergence \cite{feiguin2007interacting}, a mathematically rigorous derivation is still widely open.
The scaling limit as defined in \cite{zini2018conformal} is a double limit.  In the first step, an energy cutoff is prescribed for each system size, and a limit is taken to obtain low-energy states and a local observable algebra.  Then different system sizes are connected by matching up the low-energy states.

Our long-term goal is to develop an analogous scaling limit of the fuzzy sphere Hamiltonians \cite{he2022fuzzy} reviewed below.
Such a scaling limit is important because current numerical evidence, while highly suggestive, does not by itself establish convergence to a 3d CFT. 
The observed agreement with CFT spectra and correlators could in principle reflect finite-size effects in a limited energy window rather than a true universal continuum limit. 
To rule this out and prove convergence to a CFT, a scaling limit in a mathematically precise way in the sense of \cite{zini2018conformal} is needed. This would require analytic control over the low-energy eigenstates of the fuzzy sphere Hamiltonians. Unlike 1+1d anyon chains, fuzzy sphere Hamiltonians do not appear to be exactly solvable or have enlarged symmetries that would make such control available. We therefore leave a mathematically precise construction of the scaling limit to future work.

\section{Fuzzy sphere review}\label{sec:fuzzyreview}

Fuzzy sphere Hamiltonians exhibit 3d CFT–like spectra with only a handful of spinful fermions \cite{he2022fuzzy}. 
Their Hamiltonians are naturally written in terms of electron density operators projected to the lowest Landau level and preserve an exact $SO(3)$ symmetry, as we review here. 

\subsection{Hilbert space and fermionic density fields}

The fuzzy sphere model describes spin-$\tfrac{1}{2}$ electrons $\ppsi(\Omega) = (\psi_\uparrow(\Omega), \psi_\downarrow(\Omega))$ moving on the surface of a sphere with radius $R$ in the presence of a magnetic monopole with flux $4\pi s$ \cite{he2022fuzzy}. The monopole background quantizes the eigenstates of the system into Landau levels. When all interactions are much smaller than the Landau level spacing, we can restrict to the lowest Landau level (LLL). The LLL contains $2s+1$-fold degenerate orbitals for each spin, labeled by $m=-s,\dots,s$. The electron field $\psi_\sigma(\Omega)$ projected to the LLL may be expanded as 
\begin{align}\label{eq:fermionfieldLLL}
    \psi_\sigma(\Omega) = \frac{1}{R} \sum_{m=-s}^s Y^{(s)}_{s,m}(\Omega) \, c_{m,\sigma}, \quad \text{ where } \Omega = (\theta, \phi) \in S^2, \; \sigma \in \{\uparrow, \downarrow \}.
\end{align}
The fermion operators $c_{m,\sigma}$ obey the canonical anticommutation relations
\begin{align*}
    \{c_{m,\sigma}, c^\dagger_{m', \sigma'}\} = \delta_{\sigma,\sigma'} \delta_{m,m'}    
\end{align*}
and $Y^{(s)}_{s,m}(\Omega)$ denotes the monopole harmonics \cite{Wu:1976ge}, 
\begin{align*}
    Y^{(s)}_{s,m}(\Omega) = \sqrt{\frac{(2s+1)!}{(s+m)!(s-m)!}} e^{im\phi} \cos^{s+m}{\left( \frac{\theta}{2} \right)} \sin^{s-m}{\left( \frac{\theta}{2} \right)}.
\end{align*}
At half-filling the system contains $N=2s+1$ fermions, and the Hilbert space is spanned by states $\prod_{i=1}^N c^\dagger_{m_i, \sigma_i} \ket{0}$, where $\ket{0}$ denotes the empty vacuum state.   

The electron fields projected to the LLL \eqref{eq:fermionfieldLLL} no longer anti-commute at different locations \cite{ Fardelli:2024conformal}:
\begin{equation}\label{eq:LLLanticommutator}
\begin{aligned}
    &\{\psi_\sigma(\Omega), \psi^\dagger_{\sigma}(\Omega')\} = K(\Omega,\Omega'), \; \text{ where}\\
    &K(\Omega,\Omega') = \frac{1}{R^2} \sum_{m=-s}^s Y^{(s)}_{s,m}(\Omega) \left( Y^{(s)}_{s,m}(\Omega')  \right)^* = \frac{2s+1}{R^2} e^{2is\alpha(\Omega,\Omega')} \cos^{2s}{\left(\frac{\gamma}{2}\right)}, \text{ and }\\
    &\cos{\gamma} = \cos{\theta} \cos{\theta'} + \sin{\theta} \sin{\theta'} \cos{(\phi-\phi')}
\end{aligned}
\end{equation}
The real phase $\alpha(\Omega,\Omega') = -\alpha(\Omega', \Omega)$ is anti-symmetric as a direct consequence of $K(\Omega',\Omega)=K(\Omega,\Omega')^*$. A derivation of this relation is given in Appendix \ref{sec:Kderivation}. 

To relate the sphere radius $R$ to the monopole strength $s$, we use the magnetic field of the monopole. A monopole with charge $s$ produces a uniform radial magnetic field on $S^2$ of magnitude $B=s/R^2$. Throughout the paper, we fix $R^2=2s+1$ so that $B=\frac{s}{2s+1}$ stays finite in the thermodynamic limit $s \to \infty$.

Density operators are defined as bilinears of the fermion field,
\begin{align*}
    n^A(\Omega) = \sum_{\sigma,\lambda \in \{\uparrow, \downarrow \}} \psi^\dagger_\sigma(\Omega) A_{\sigma \lambda} \, \psi_\lambda(\Omega),
\end{align*}
where $A$ is a $2\times2$ matrix acting on the spin.
In particular, we will make frequent use of 
\begin{align*}
    n^0(\Omega) = \sum_{\lambda=\uparrow,\downarrow} \psi_\lambda^\dagger(\Omega) \psi_\lambda(\Omega) \quad \text{and} \quad n^{\alpha}(\Omega) = \sum_{\lambda,\eta \in \{\uparrow, \downarrow\}} \psi^{\dagger}_\lambda(\Omega) \,(\sigma^\alpha)_{\lambda\eta} \,  \psi_\eta(\Omega), \quad \alpha=x,y,z,
\end{align*}
with $\sigma^\alpha$ denoting the Pauli matrices. The $n^A(\Omega)$ fields can be expanded in spherical harmonics and the corresponding density modes $n^A_{l,m}$ defined by
\begin{align*}
    n^A(\Omega) &= \frac{1}{2s+1} \sum_{l=0}^{2s} \sum_{m=-l}^l n^A_{l,m} Y_{l,m}(\Omega), \quad  n^A_{l,m} = (2s+1) \int d\Omega \; Y^*_{l,m}(\Omega) n^A(\Omega).
\end{align*}
After writing $\psi_\sigma(\Omega)$ in the orbital basis, these mode operators admit the explicit bilinear form
\begin{equation}\label{eq:densitymodes}
\begin{aligned} 
    n^A_{l,m} &=(2s+1) \sqrt{2l+1} \begin{pmatrix}
        s & s & l \\ s & -s & 0
    \end{pmatrix} \sum_{m_1=-s}^s (-1)^{s+m_1} \begin{pmatrix}
        s & s & l \\ m_1 & m-m_1 & -m
    \end{pmatrix} \sum_{\sigma, \lambda }c^\dagger_{m_1,\sigma} c_{m_1-m,\lambda} A_{\sigma \lambda}
\end{aligned}
\end{equation}
with $Y_{l,m}(\Omega)$ denoting spherical harmonics and $\int d\Omega = \int d\phi \, d\theta \sin{\theta}$. 

\subsection{Hamiltonian and symmetries}

The fuzzy sphere Hamiltonian that realizes the 3d Ising CFT is given by \cite{he2022fuzzy, Hofmann2023qmc}
\begin{align}\label{eq:Hamfields}
    H = \int d\Omega \left( U_0 \left( n^0(\Omega)^2 - n^z(\Omega)^2 \right) + U_1 \left( n^0(\Omega) \nabla^2 n^0(\Omega) - n^z(\Omega) \nabla^2 n^z(\Omega) \right) - h n^x(\Omega) \right)
\end{align}
Expanding in the density modes $n^A_{l,m}$, this becomes
\begin{align}\label{eq:Hamdensitymodes}
    H = -h n^x_{0,0} + \sum_{l=0}^{2s} \sum_{m=-l}^l (-1)^m \left(U_0 + U_1 \frac{l(l+1)}{2s+1}\right) \left( n^0_{l,m} n^0_{l,-m} - n^z_{l,m} n^z_{l,-m} \right),
\end{align}
where we used the hermiticity property
\begin{align*}
    \left( n^A_{l,m} \right)^\dagger = (-1)^m n^{A^\dagger}_{l,-m}.
\end{align*}
The filling (total number of electrons) is
\begin{align*}
    N_{\text{el}} = \sum_{m=-s}^s (n^\uparrow_m + n^\downarrow_m) = (-1)^{2s} n^0_{0,0},
\end{align*}
and commutes with every $n^A_{l,m}$. To study the 3d Ising CFT in the fuzzy sphere model, one can restrict to the sector with filling $N_\text{el}=N$.
For large $h$, the model is in the trivial phase with a unique ground state. For small $h$, it enters a quantum Hall ferromagnet phase with two degenerate ground states, spontaneously breaking the $\mathbb{Z}_2$ spin-flip symmetry acting as $c_{m,\uparrow} \leftrightarrow c_{m,\downarrow}$. In the $(h,U_0)$ phase diagram (with $U_1$ fixed), these phases are separated by a critical line along which the low-energy spectrum organizes into conformal multiplets consistent with the 3d Ising CFT. Because finite-size effects vary appreciably along this line, it is useful in practice to tune both $h$ and $U_1$.

Crucially, the Hamiltonian preserves an exact $SO(3)$ rotational symmetry, which appears to be important for extracting CFT data already at small system sizes. The $SO(3)$ generators act on the orbital indices in the spin-$s$ representation and are proportional to the $n^0_{1m}$ density modes:
\begin{equation}\label{eq:so3generators}
    \begin{aligned}
        J_z &= (s+1) \frac{(-1)^{2s}}{\sqrt{3}} n^0_{1,0} = \sum_{m=-s}^s m \left( c^\dagger_{m,\uparrow} c_{m,\uparrow} + c^\dagger_{m,\downarrow} c_{m,\downarrow} \right), \\
        J_\pm &= \pm(s+1) (-1)^{2s} \sqrt{\frac{2}{3}} n^0_{1,\pm 1} = \sum_{m=-s}^s \sqrt{(s\mp m)(s \pm m+1)} \left( c^\dagger_{m\pm1,\uparrow} c_{m,\uparrow} + c^\dagger_{m\pm 1,\downarrow} c_{m,\downarrow} \right).   
    \end{aligned}
\end{equation}
Using the commutation relations \eqref{eq:modescommutator} reviewed below, it is straightforward to check that $[H,J_z]=[H,J_\pm]=0$. Since spatial rotations only act on the orbital sector $V^s \cong \mathbb{C}^{2s+1}$ (not on the internal spin/flavor indices), the relevant representation theory is that of operators on $V^s$, i.e. $\mathrm{End}(V_s)$. Under the adjoint $so(3)$ action, $\mathrm{End}(V_s)$ decomposes as 
\begin{align*}
    \mathrm{End}(V_s) \cong V^s \otimes \overline{V^s} = \oplus_{l=0}^{2s} \; V^l.
\end{align*}
Accordingly, any rotation-invariant fuzzy sphere Hamiltonian has eigenstates that organize into $SO(3)$ multiplets labeled by $l$.

Beyond the Ising CFT, similar fuzzy sphere Hamiltonians have been used to realize a variety of 3d CFTs, including the free real scalar CFT \cite{He2025scalar, Taylor:2025odf}, the $O(N)$ and Wilson-Fisher CFTs \cite{Dey:2025zgn, Guo:2025odn}, DQCPs with emergent $SO(5)$ symmetry \cite{PhysRevX.14.021044, PhysRevB.110.125153}, the non-unitary Yang-Lee CFT \cite{Fan:2025bhc, EliasMiro:2025msj, ArguelloCruz:2025zuq}, a superconformal Ising CFT \cite{Tang:2025wtj}, and the free fermion \cite{Zhou:2025fermion, voinea2026majorana} and Chern-Simons-matter CFTs \cite{Zhou:2024zud, Zhou:2025chernsimons}. In addition, fractional-filling fuzzy sphere models (rather than integer $\nu=1$ filling) can also realize 3d CFTs \cite{voinea2025anyons}. 
These examples suggest broad applicability, but a general characterization of which 3d CFTs can be realized by fuzzy sphere Hamiltonians remains an open problem.

\subsection{Conformal generators for the fuzzy sphere Hamiltonian}\label{subsec:conformalgenerators}

In radial quantization on \(S^2\times \mathbb R\), the dilatation operator $D$ is identified with the Hamiltonian $H$, up to an additive constant and an overall normalization.  
Refs.~\cite{Fardelli:2024conformal} and \cite{Fan:2024vcz} found lattice precursors of the conformal generators $\Lambda_m = K_m + P_m$ with $m \in \{0, \pm 1\}$ ($\Lambda_0=\Lambda_z$), where $K_m$ and $P_m$ generate translations and special conformal transformations as in \eqref{eq:confgenerators}.
In the continuum CFT on \(S^2\times \mathbb R\), this combination is obtained from the \(\ell=1\) moment of the stress tensor component \(T_{00}\). On the fuzzy sphere, the corresponding lattice operator is defined using the microscopic Hamiltonian density \(H(\Omega)\) in \eqref{eq:Hamfields}:
\begin{align*}
    \Lambda_m = \int d\Omega \; Y_{1,m}(\Omega)\, H(\Omega).
\end{align*}
Together with the identification of the dilatation operator $D$ with the Hamiltonian $H$, this gives lattice representatives of $P_m$ and $K_m$ through the conformal algebra relation $[D,P_m+K_m]=P_m-K_m$, namely
\[
P_m=\frac12\bigl(\Lambda_m+[D,\Lambda_m]\bigr),
\qquad
K_m=\frac12\bigl(\Lambda_m-[D,\Lambda_m]\bigr).
\]
Because $H(\Omega)$ is invariant under $\mathbb{Z}_2$ spin-flip, these lattice precursors inherit the same symmetry. 

At finite $s$, this identification does not satisfy the $so(3,2)$ algebra exactly. For example, the commutator $[K_z, P_z] = 2D$ (equivalently, $[\Lambda_z,[D,\Lambda_z]] = 4D$) holds only within a low-energy subspace, as demonstrated numerically in \cite{Fardelli:2024conformal}. The energy window over which this relation is well satisfied grows with $\sqrt{s}$, consistent with conformal symmetry emerging in the continuum limit.

The overlap of the lattice precursors with the CFT generators can be improved by tuning couplings ($h$, $U_0$, $U_1$) and adding irrelevant lattice terms to the Hamiltonian density used in defining $\Lambda_m$. In that sense, the identification of lattice precursors is not unique. 

\section{Fuzzy sphere algebra}\label{sec:fuzzyspherealgebra}

The angular momentum modes of the electron density obey an interesting algebra that can be viewed as a spin-enriched version of the Girvin-MacDonald-Platzman (GMP) algebra on the sphere \cite{He2025scalar}. 
Section~\ref{subsec:commrel} reviews this algebra and collects results that use its concrete fermion-bilinear realization; Section~\ref{subsec:jacobicommuting} treats the commutator relations as defining an abstract bracket and proves representation-independent properties.

\subsection{Commutation relations of the density modes}\label{subsec:commrel}

The mode operators \eqref{eq:densitymodes} satisfy the following commutation relations \cite{He2025scalar}: 
\begin{equation}\label{eq:modescommutator}
    \begin{aligned}
        \left[ n^A_{l,m}, n^B_{l',m'} \right] = &\sum_{\substack{L = 0 \\ l+l'+L \text{ even}}} h_L(s,l,l') (-1)^{m+m'} \begin{pmatrix}
            l & l' & L \\ m & m' & -m-m'
        \end{pmatrix}  n^{[A,B]}_{L,m+m'} \\
        - &\sum_{\substack{L = 0 \\ l+l'+L \text{ odd}}} h_L(s,l,l') (-1)^{m+m'} \begin{pmatrix}
            l & l' & L \\ m & m' & -m-m'
        \end{pmatrix}  n^{\{A,B\}}_{L,m+m'},
    \end{aligned}
\end{equation}
where 
\begin{align}
    h_L(s,l,l') = (2s+1) \sqrt{(2l+1)(2l'+1)(2L+1)} \begin{Bmatrix}
        l & l' & L \\ s & s & s
    \end{Bmatrix}
    \frac{\begin{pmatrix}
        s & s & l \\ s & -s & 0
    \end{pmatrix} \begin{pmatrix}
        s & s & l' \\ s & -s & 0
    \end{pmatrix}}{
    \begin{pmatrix}
        s & s & L \\ s & -s & 0
    \end{pmatrix}
    }.
\end{align}
We derive the density–mode commutator in Appendix~\ref{sec:commutatorderivation} by rewriting the mode operators in a basis of fuzzy spherical harmonics $\hat{Y}_{l,m}$ \cite{Madore:1991bw,Grosse:1995ar,Iso:2001mg}. The matrices $\hat{Y}_{l,m}$ furnish a $SO(3)$ covariant basis of the matrix algebra $\text{Mat}_{2s+1}$. Their matrix entries are given by
\begin{align*}
    \left(\hat{Y}_{l,m}\right)_{ab} = (-1)^{s-a} \sqrt{2l+1} \begin{pmatrix}
        s & s & l \\ 
        a & -b & -m
    \end{pmatrix},
\end{align*}
and they satisfy
\begin{align*}
    (\hat{Y}_{l,m})^\dagger = (-1)^m \hat{Y}_{l,-m} \quad \text{and } \Tr{(\hat{Y}_{l,m})^\dagger \hat{Y}_{l',m'}} = \delta_{l,l'} \delta_{m,m'}.
\end{align*}
Up to a prefactor, the density modes can be written as fermion bilinears with orbital matrix $\hat{Y}_{l,m}$:
\begin{equation}\label{eq:densitymodesTlm}
\begin{aligned}
    &n^A_{l,m} = \mathcal{N}_{s,l} \sum_{m_1, m_2=-s}^s \left(\hat{Y}_{l,m}\right)_{m_1,m_2} c^\dagger_{m_1,\sigma} A_{\sigma\lambda}c_{m_2,\lambda}, \quad \text{with } \mathcal{N}_{s,l} = (2s+1) \begin{pmatrix} s & s & l \\ s & -s & 0\end{pmatrix}.
\end{aligned}
\end{equation}
We suppress explicit summation over the spin indices in what follows.

The fermion bilinears $E_{ij} := c^\dagger_i c_j$ with combined orbital-spin indices $i=(m_1,\sigma)$, $j=(m_2,\lambda)$ generate $u(2N)$ ($N=2s+1$) with the standard commutation relations 
\begin{align*}
    \left[E_{ij}, E_{kl} \right] = \delta_{jk} E_{il} - \delta_{li} E_{kj}.
\end{align*}
In their fermionic representation \eqref{eq:densitymodesTlm}, each mode $n^A_{l,m}$ is a linear combination of $E_{ij}$. Using the completeness of the orbital and spin bases, this change of basis is invertible; thus, the two sets span the same vector space and generate the same matrix Lie algebra $u(2N)$.

Besides the commutation relations, there are additional polynomial relations among the fermionic density operators. Using the LLL anticommutation relation, the squared density field can be written
as \cite{He2025scalar}
\begin{equation}
\begin{aligned}
    n^A(\Omega)^2 &= \sum_{\alpha \beta \gamma \delta} \psi^\dagger_\alpha(\Omega) A_{\alpha \beta} \psi_\beta(\Omega) \psi^\dagger_\gamma(\Omega) A_{\gamma \delta} \psi_\delta(\Omega) \\
    &= \sum_{\alpha \beta \gamma \delta} A_{\alpha \beta} A_{\gamma \delta} \left(\psi^\dagger_\alpha(\Omega) K(\Omega, \Omega) \delta_{\beta \gamma} \psi_\delta(\Omega) - \psi^\dagger_\alpha(\Omega) \psi^\dagger_\gamma(\Omega) \psi_\beta(\Omega) \psi_\delta(\Omega) \right) \\
    &= n^{A^2}(\Omega) - \sum_{\alpha \beta \gamma \delta} A_{\alpha \beta} A_{\gamma \delta} \psi^\dagger_\alpha(\Omega) \psi^\dagger_\gamma(\Omega) \psi_\beta(\Omega) \psi_\delta(\Omega) 
\end{aligned}
\end{equation}
For $A\in\{\sigma^x,\sigma^y,\sigma^z\}$ we have $A^2=\mathbb I$, and the quartic term can be eliminated
in favor of $n^0(\Omega)^2$ using
\begin{equation*}
    n^0(\Omega)^2 = n^0(\Omega) - \sum_{\alpha,\beta}
    \psi^\dagger_\alpha(\Omega)\psi^\dagger_\beta(\Omega)\psi_\alpha(\Omega)\psi_\beta(\Omega),
\end{equation*}
which yields
\begin{equation*}
    n^x(\Omega)^2=n^y(\Omega)^2=n^z(\Omega)^2 = 2n^0(\Omega) - n^0(\Omega)^2.
\end{equation*}
Integrating over $\Omega$ gives the corresponding quadratic relations among density modes:
\begin{equation}\label{relations: fuzzy}
    \sum_{l,m} (-1)^m n^\alpha_{l,m}\,n^\alpha_{l,-m}
    = 2(2s+1)n^0_{0,0} - \sum_{l,m} (-1)^m n^0_{l,m}\,n^0_{l,-m},
    \qquad \alpha=x,y,z.
\end{equation}

\subsection{Jacobi identity and commuting subalgebras}\label{subsec:jacobicommuting}

Does the mode algebra, viewed abstractly through the commutation relations \eqref{eq:modescommutator}, define a genuine Lie algebra? Let $\mathcal{A}_s = \{\Tilde{n}^A_{l,m}\}_{0 \leq l=0 \leq 2s, |m|\leq l}$ be the algebra spanned by formal generators $\Tilde{n}^A_{l,m}$, equipped with the bracket defined by \eqref{eq:modescommutator}. 
A priori this bracket need not satisfy Jacobi.
To verify Jacobi, we use the fermionic realization. Let $\rho: \mcA_s \to \End(\mathcal H_{\rm Fock})$ be the map that sends the abstract generators to the corresponding lowest-Landau-level bilinears,
\[
\rho(\Tilde{n}^A_{l,m}) := n^A_{l,m}
\qquad A\in\{\mathbb I,\sigma^x,\sigma^y,\sigma^z\},
\]
where $n^A_{l,m}$ on the right-hand side is the explicit operator defined in \eqref{eq:densitymodes}.
Since $\End(\mathcal H_{\rm Fock})$ is an associative algebra, its commutator bracket satisfies Jacobi.
Hence for all $x,y,z\in \mcA_s$ the Jacobiator
\[
J(x,y,z):=[x,[y,z]]+[y,[z,x]]+[z,[x,y]]
\]
obeys $\rho(J(x,y,z))=0$.
Therefore, if $\rho$ is faithful, it follows that $J(x,y,z)=0$ already in $\mcA_s$,
i.e.\ the commutator \eqref{eq:modescommutator} defines a genuine Lie algebra.

We prove faithfulness by showing that the representation matrices $\{n^A_{l,m}\}$ are linearly independent. If the linear relation $\sum_{i} \alpha_{i} n^{A_i}_{l_i,m_i} = 0$ holds, then $\sum_{i} \alpha_{i} [n^{A_i}_{l_i,m_i}, c^\dagger_{r,\eta}] = 0$ also holds. Inserting the fermionic bilinear form \eqref{eq:densitymodesTlm} into the latter equation and using
\begin{align*}
    [c^\dagger_{m_1,\sigma} c_{m_2,\lambda}, c^\dagger_{r,\eta}] = \delta_{r,m_2} \delta_{\lambda,\eta} c^\dagger_{m_1,\sigma}
\end{align*}
results in 
\begin{align*}
    \sum_{i} \alpha_{i} \left[n^{A_i}_{l_i,m_i}), c^\dagger_{r,\eta}\right] = \sum_{i} \alpha_{i} \,\mathcal{N}_{s,l_i} \sum_{m_1} \left(\hat{Y}_{l_i,m_i}\right)_{m_1,r} A^i_{\sigma_i,\eta} c^\dagger_{m_1,\sigma_i} = 0.
\end{align*}
The creation operators $c^\dagger_{m_1,\sigma_i}$ are linearly independent, requiring the expression above to vanish for each $m_1,r,\sigma,\eta$. Because the $A_i$ form a basis of Mat$_2$ and the fuzzy spherical harmonics $\hat{Y}_{l,m}$ a basis of Mat$_{2s+1}$, this implies $\alpha_i = 0$ for all $i$. Therefore, the $\{n^A_{l,m}\}$ are linearly independent and $\rho$ is faithful. 

It would be very interesting to characterize this mode algebra more intrinsically, i.e. to find a presentation of each fuzzy sphere algebra $\mathcal{A}_s$ with generators $\{\Tilde{n}^A_{l,m}\}$ and a minimal set of relations that define the algebra $\mathcal{A}_s$, potentially including identities between squared modes as in \eqref{relations: fuzzy}.
If one can endow $\mcA_s$ with an associative multiplication $x \cdot y$ such that, for the generators, the induced commutator  $[x,y]=x\cdot y-y\cdot x$ agrees with the bracket in \eqref{eq:modescommutator}, then the Jacobi identity would follow automatically: the commutator in any associative algebra is a Lie bracket.

The commutator \eqref{eq:modescommutator} comes with strong angular-momentum selection rules, and these imply the existence of abelian subalgebras at the level of the abstract algebra. In the spinless sector spanned by $\{\Tilde{n}^0_{l,m}\}$, one finds three families of mutually commuting modes \cite{dora2022gmp}:
\begin{align*}
    \left[\Tilde{n}^0_{l,0}, \Tilde{n}^0_{l',0}\right] = 0, \quad \left[\Tilde{n}^0_{l,l}, \Tilde{n}^0_{l',l'}\right] = 0, \quad \left[\Tilde{n}^0_{l,-l}, \Tilde{n}^0_{l',-l'}\right] = 0.
\end{align*}
The first commutator vanishes because the 3j symbol $\begin{pmatrix}
    l & l' & L \\ 0 & 0 & 0
\end{pmatrix}$ that appears in \eqref{eq:modescommutator} vanishes for $l+l'+L$ odd. The second commutator can only contain modes with $m=l+l' \leq L$, but the selection rules of the 3j symbol require $L \leq l+l'$, so the only possibility is $L=l+l'$. This choice makes $l+l'+L$ even, and so the term is excluded by the $l+l'+L$ odd constraint in the sum. A similar argument shows that the third commutator vanishes.
Therefore, the rank of the subalgebra generated by the $\Tilde{n}^0_{l,m}$ is $N=2s+1$. For the full algebra generated by the $\Tilde{n}^A_{l,m}$, the rank doubles to $2N$, and a natural maximal commuting subalgebra is generated by $\{\Tilde{n}^0_{l,0},\,\Tilde{n}^z_{l,0}\}$.

Finally, a separate question is whether this algebra admits a nontrivial central extension; we briefly comment on this in Appendix~\ref{sec:centralextension}.

\section{Thermodynamic limit of the density mode algebra}\label{sec:thermodymlimit}

    The fuzzy sphere geometry realized by these lowest Landau level models is encoded in the projected coordinate matrices
    \begin{align*}
        (X_i)_{m_1,m_2}
        =
        \int d\Omega \; x_i(\Omega)\,
        \frac{\bigl(Y^{(s)}_{s,m_1}(\Omega)\bigr)^*}{R}
        \frac{Y^{(s)}_{s,m_2}(\Omega)}{R},
        \qquad m_1,m_2=-s,\ldots,s .
    \end{align*}
    They satisfy
    \begin{align*}
        [X_i,X_j]
        =
        \frac{iR}{s+1}\,\epsilon_{ijk}X_k,
        \qquad
        \sum_i X_i^2
        =
        \frac{s}{s+1}R^2\,\mathbb I .
    \end{align*}
    This geometry has two fundamentally different large-$s$ limits \cite{Chu:2001xi, Steinacker_book}: First, one can zoom into a microscopic patch near the north pole. There $X_3\simeq R$, and the remaining coordinates obey the fuzzy plane commutation relation
    \begin{align*}
        [X_1, X_2] = i \Theta, \qquad \Theta = \frac{R^2}{s+1},
    \end{align*}
    with $\Theta$ remaining finite in the $s \to \infty$ limit.

    Second, one can take the large-$s$ limit globally, without zooming into a local patch. Since $R/(s+1)\to0$ for $R\sim\sqrt{s}$, the coordinate commutators vanish and the fuzzy sphere approaches an ordinary commutative sphere.

    In Section~\ref{subsec:planarlimit}, we show that in the local planar limit, the density modes with large angular momentum $l \sim \sqrt{s}$ become generators of the GMP algebra. 
    In Section~\ref{subsec:commutinglimit}, we turn to the global commutative limit, where the small-angular-momentum modes $l\ll\sqrt{s}$ obey semiclassical commutation relations. Finally, in Section~\ref{subsec:oscillator}, we further restrict to a low-excitation sector above the paramagnetic reference state and show that the density modes reduce to approximate harmonic oscillators. The restriction to such a low-energy subspace is believed to be necessary to see the emergence of conformal symmetry.

\subsection{Quantum planar limit of \texorpdfstring{large-$l$}{large-l} density modes}\label{subsec:planarlimit}

We first consider the local planar limit of the fuzzy sphere geometry. In this limit \(s\to\infty\) while one zooms into a microscopic patch near the north pole, and the relevant spherical modes have angular momentum \(l\sim\sqrt{s}\), corresponding to finite planar momentum \(q\sim l/R\). The fuzzy spherical harmonics \(\hat Y_{l,m}\) entering the density modes \eqref{eq:densitymodesTlm} can then be approximated using the large-\(s\) asymptotics of Clebsch--Gordan coefficients \cite[\S 8.9.1]{Varshalovich:1988ifq}, following \cite[Appendix E]{Matsuura:2015caa}.
\begin{align}\label{eq:Tlm_to_Ylm}
    \left(\hat{Y}_{l,m}\right)_{m_1,m_2} = (-1)^{s-m_2}  \,C^{lm}_{s m_1,s (-m_2)}
    \ \approx\
    \delta_{m,m_1-m_2}\,\sqrt{\frac{4\pi}{2s+1}}\,Y_{l m}(\theta,0),
\end{align}
where the small polar angle \(\theta\) is related to the row index $m_1$ by 
\begin{equation}\label{eq:theta_def}
    \sin\frac{\theta}{2}=\sqrt{\frac{m_1+\tfrac12}{2s+1}}.
\end{equation}
(Here the north-pole patch corresponds to \(m_1\ll s\), so \(\theta\ll 1\).)
When $l \sim \sqrt{s}$, one can further use the Bessel function approximation of spherical harmonics in the limit
\[
l\to\infty,\quad m\to\infty,\quad \theta\to 0,\quad\text{with } l\theta\ \text{fixed},
\]
which yields \cite[\S 5.12.3]{Varshalovich:1988ifq}
\begin{equation}\label{eq:Ylm_to_Bessel}
    Y_{l m}(\theta,0)\ \approx\ \sqrt{\frac{l}{2\pi}}\,(-1)^m\,J_m(l\theta).
\end{equation}
The resulting approximation for the fuzzy spherical harmonics is 
\begin{align}\label{eq:Tlm_planar_asymptotics}
    \left(\hat{Y}_{l,m}\right)_{m_1,m_2}\ \approx \delta_{m,m_1-m_2}\,\sqrt{\dfrac{2l}{2s+1}}\,(-1)^m\,J_m(l\theta).
\end{align}

Next, it is useful to rewrite \eqref{eq:Tlm_planar_asymptotics} in a form that makes the connection to the noncommutative plane clearer \cite{Matsuura:2015caa}, 
\begin{align}\label{eq:Tlm_as_plane_wave}
    \left(\hat{Y}_{l,m}\right)_{m_1,m_2}
    \ \approx\ 
    (-i)^m\,\sqrt{\frac{2l}{2s+1}}\int_{0}^{2\pi}\frac{d\phi}{2\pi}\,e^{im\phi}
    \left(e^{\,i\mathbf q(\phi)\cdot {\mathbf X}}\right)_{m_1,m_2},
    \qquad |\mathbf q(\phi)|=\frac{l}{R},
\end{align}
where \(e^{\,i\mathbf q\cdot {\mathbf X}}\) is the plane-wave operator on the quantum plane, with noncommuting coordinates
\begin{equation}\label{eq:Moyal_comm}
    [X_1, X_2]=i\Theta,\qquad \Theta = \frac{R^2}{s+1}\sim \frac{1}{2B}.
\end{equation}
Inserting \eqref{eq:Tlm_as_plane_wave} into the definition of the density modes \eqref{eq:densitymodesTlm} gives
\begin{align}\label{eq:nlm_to_rhoq}
    n^A_{l,m}\ \approx\ 
    \kappa_{s,l}\int_{0}^{2\pi}\frac{d\phi}{2\pi}\,e^{im\phi}\,\rho^A_{\mathbf q(\phi)},
    \qquad |\mathbf q(\phi)|=\frac{l}{R},
\end{align}
with some prefactor \(\kappa_{s,l}\), and where the planar density modes are
\begin{align}\label{eq:rhoq_def}
    \rho^A_{\mathbf q}
    :=\sum_{m_1,m_2= 0}^\infty
    \left(e^{\,i\mathbf q\cdot \hat{\mathbf x}}\right)_{m_1,m_2}\, c^\dagger_{m_1,\sigma}\,A_{\sigma\lambda}\,c_{m_2,\lambda}.
\end{align}
Thus the large-\(l\) spherical density modes become angular Fourier components of planar momentum modes \(\rho^A_{\mathbf q}\) with \(|\mathbf q|=l/R\).
Density modes with small angular momentum $l \ll \sqrt{s}$ collapse to $\mathbf{q}=0$ modes on the plane (which commute with finite momentum modes when $[A,B]=0$).

The operators \(e^{i\mathbf q\cdot{\mathbf X}}\) satisfy the algebra following from \eqref{eq:Moyal_comm},
\begin{equation*}
    e^{i\mathbf q\cdot {\mathbf X}}\,e^{i\mathbf p\cdot {\mathbf X}}
    =\exp\!\Big(-\frac{i}{2}\Theta\,(\mathbf q\times \mathbf p)\Big)\,
    e^{i(\mathbf q+\mathbf p)\cdot {\mathbf X}},
    \qquad \mathbf q\times\mathbf p:=q_1p_2-q_2p_1.
\end{equation*}
Using the fermion bilinear expansion \eqref{eq:rhoq_def}, this implies the spin-enriched planar GMP algebra
\begin{align}\label{eq:GMP_spin}
    \left[\rho^A_{\mathbf q},\rho^B_{\mathbf p}\right]
    =\cos\!\Big(\frac{\Theta}{2}\,(\mathbf q\times \mathbf p)\Big)\,
    \rho^{[A,B]}_{\mathbf q+\mathbf p}
    +i\sin\!\Big(\frac{\Theta}{2}\,(\mathbf q\times \mathbf p)\Big)\,
    \rho^{\{A,B\}}_{\mathbf q+\mathbf p}
\end{align}
The spinless subalgebra ($A=B=\mathbb{I}$) reduces to the planar GMP algebra \cite{girvin1985collective}.

From \eqref{eq:GMP_spin}, the commutator of the $l \sim \sqrt{s}$ modes in the large-$s$ limit can be written as an angular average of the planar GMP commutator:
\begin{align}\label{eq:nAlm_comm_limit}
    \big[ n^A_{l,m}, n^B_{l',m'} \big]
    \;\approx\; \kappa_{s,l} \kappa_{s,l'}
    \int_0^{2\pi}\frac{d\phi}{2\pi}\int_0^{2\pi}\frac{d\phi'}{2\pi}\;
    e^{im\phi}e^{im'\phi'}\,
    \Big(
        \cos\beta\,\rho^{[A,B]}_{\mathbf q(\phi)+\mathbf q'(\phi')}
        - i\sin\beta\,\rho^{\{A,B\}}_{\mathbf q(\phi)+\mathbf q'(\phi')}
    \Big),
\end{align}
where
\begin{equation*}
    \mathbf q(\phi)=\frac{l}{R}(\cos\phi,\sin\phi),\qquad
    \mathbf q'(\phi')=\frac{l'}{R}(\cos\phi',\sin\phi'),\qquad
    \beta=\frac{\Theta}{2}\big(\mathbf q(\phi)\times \mathbf q'(\phi')\big).
\end{equation*}
It is convenient to change variables to $\Phi=\frac{\phi+\phi'}{2}$ and $\Delta=\phi'-\phi$
so that
\begin{equation*}
    e^{im\phi}e^{im'\phi'}=e^{i(m+m')\Phi}\,e^{i(m'-m)\Delta/2}.
\end{equation*}
In these variables, the scalars \(\beta\) and the magnitude
\(
|\mathbf Q|:=|\mathbf q(\phi)+\mathbf q'(\phi')|
\)
depend only on \(\Delta\) (through \(\sin\Delta\) and \(\cos\Delta\)), whereas the direction of
\(
\mathbf Q:=\mathbf q(\phi)+\mathbf q'(\phi')
\)
is
\begin{equation*}
    \varphi_{\mathbf Q}=\Phi+\alpha(\Delta), 
    \quad \text{with } \alpha(\Delta) =\mathrm{arctan2}\left((l'-l) \sin{\Delta/2}, (l+l')\cos{\Delta/2} \right).
\end{equation*}
Writing the inverse angular decomposition (the inverse of \eqref{eq:nlm_to_rhoq})
\begin{equation}\label{eq:rho_inverse}
    \rho^A_{\mathbf Q}
    \;=\;
    \sum_{M\in\mathbb Z} e^{-iM\varphi_{\mathbf Q}}\; n^A_{L,M},
    \qquad
    L:=|\mathbf Q|\,R,
\end{equation}
and inserting \eqref{eq:rho_inverse} into \eqref{eq:nAlm_comm_limit}, the integral over \(\Phi\) enforces angular-momentum conservation:
\begin{equation*}
    \int_0^{2\pi}\frac{d\Phi}{2\pi}\,e^{i(m+m'-M)\Phi}=\delta_{M,m+m'}.
\end{equation*}
As a result,
\begin{align*}
    \big[ n^A_{l,m}, n^B_{l',m'} \big] \approx
    \int_0^{2\pi}\frac{d\Delta}{2\pi}\;
    e^{i(m'-m)\Delta/2}\,e^{-i(m+m')\alpha(\Delta)}
    \Big(
        \cos\beta(\Delta)\,n^{[A,B]}_{L(\Delta),\,m+m'}
        - i\sin\beta(\Delta)\,n^{\{A,B\}}_{L(\Delta),\,m+m'}
    \Big),
\end{align*}
where \(L(\Delta)=|\mathbf q+\mathbf q'|\,R=\sqrt{l^2+l'^2+2ll'\cos\Delta}\) and \(\beta(\Delta)=\frac{\Theta}{2}\frac{ll'}{R^2}\sin\Delta\).

We can use the large-$l$ correspondence \eqref{eq:nlm_to_rhoq} to rewrite the fuzzy sphere Hamiltonian \eqref{eq:Hamdensitymodes} in the planar limit. Substituting $n^A_{l,m}\mapsto \rho^A_{\mathbf q}$ (with $|\mathbf q|=l/R$) yields
\begin{align}\label{eq:ham_limit}
    H \approx -h \rho^x_{\mathbf 0} + \int d^2q \; V(q)\Big( \rho^0_{\mathbf q}\rho^0_{-\mathbf q} - \rho^z_{\mathbf q}\rho^z_{-\mathbf q}\Big),
    \qquad V(q)=U_0+U_1 q^2.
\end{align}
The first term acts as a uniform field, and the remaining terms describe planar density--density interactions.

\subsection{Semiclassical limit of \texorpdfstring{small-$l$}{small-l} density modes}\label{subsec:commutinglimit}

We next consider the global commutative limit of the fuzzy sphere geometry. Here \(s\to\infty\) is taken on the full sphere, without zooming into a local patch, and we focus on density modes with small angular momentum \(l\ll\sqrt{s}\). In this regime the noncommutativity scale vanishes, so the fuzzy sphere approaches an ordinary sphere and the density mode algebra becomes semiclassical. 

In the large-$s$ limit, the fuzzy spherical harmonics matrices \(\hat Y_{l,m}\) with \(l\ll \sqrt{s}\) can be identified with the ordinary spherical harmonics \(Y_{l,m}(\Omega)\) on \(S^2\) \cite{Steinacker_book}. Moreover, the matrix product approaches the pointwise product of functions, while the commutator reduces to the Poisson bracket on the sphere. Therefore,
\begin{align}\label{eq:comm_limit_commutative}
     \big[ n^A[f], n^B[g] \big]
    \;=\;
    n^{[A,B]}[f g]
    \;+\;
    \mcO\!\left(\frac{1}{s}\right)\, i\, n^{\{A,B\}}\!\left[\{f,g\}_{\rm PB}\right],
\end{align}
where $n^A[f]$ is defined as 
\begin{align*}
    n^A[f] = \int d\Omega \; f(\Omega) \psi^\dagger_\sigma(\Omega) A_{\sigma \lambda} \psi_\lambda(\Omega).
\end{align*}

Taking $f=Y_{l,m}$ and $g=Y_{l',m'}$ with $l,l'\ll \sqrt{s}$ yields the semiclassical commutator of the corresponding small-$l$ density modes. Using the standard product decomposition of spherical harmonics, the leading term is
\begin{equation}\label{eq:comm_semiclassical}
\begin{aligned}
&[n^A_{l,m},n^B_{l',m'}]
\;\approx\; n^{[A,B]}\!\left[\,Y_{l,m}Y_{l',m'}\,\right] \\
&=
\sum_{L}
\sqrt{(2l+1)(2l'+1)(2L+1)}\;
\begin{pmatrix}
l & l' & L \\ 0 & 0 & 0
\end{pmatrix}
\begin{pmatrix}
l & l' & L \\ m & m' & -(m+m')
\end{pmatrix}
(-1)^{m+m'}\,
n^{[A,B]}_{L,\,m+m'} .
\end{aligned}
\end{equation}
Here the first $3j$ symbol enforces the triangle inequalities (in particular $L \leq l+l'$, so the result remains in the small-$l$ sector) and is nonzero only when $l+l'+L$ is even. 

It is important, however, that the semiclassical bracket is a limit of the full density mode algebra; it does not include any projection to the low-energy subspace of a particular Hamiltonian. For this reason, it should not be expected by itself to reproduce relations that emerge only in the low-energy sector of critical fuzzy sphere Hamiltonians. For example, it does not directly imply the conformal commutation relation $[\Lambda_z,[D,\Lambda_z]]=4D$ observed numerically in \cite{Fardelli:2024conformal}, where $D$ is identified with $H$ and $\Lambda_z$ with the $l = 1$ moment of the Hamiltonian density. Capturing such emergent conformal relations requires both the large-$s$ limit and a restriction to low-energy states, and therefore likely requires a more refined scaling limit of the type discussed in Sec.~\ref{subsec:scalinglimit}.

\subsection{Oscillator approximation in subspace above paramagnetic state}\label{subsec:oscillator}

The small-$l$ density mode algebra simplifies further when restricted to a low-excitation sector.
In \cite{He2025scalar}, it was observed that, in the subspace generated by a few small angular momentum spin flips above the paramagnetic reference state (which we here write in the $z$-basis)
\(\ket{\Downarrow}:=\ket{\downarrow\downarrow\cdots\downarrow}\),
the density modes \(n^+_{l,m}\) and \(n^-_{l,m}\) behave approximately as independent harmonic oscillators. 
A similar Bogoliubov-rotated oscillator description also appears to govern the low-energy sector of the critical fuzzy sphere model for the free scalar CFT. In this Section, we make the oscillator approximation more precise by bounding operator norms of terms in the commutator.

The commutators \([n^+_{l,m},n^+_{l',m'}]=[n^-_{l,m},n^-_{l',m'}]=0\) hold exactly, while the mixed commutator is given by
\begin{equation}\label{eq:commutatornplusminus}
    \begin{aligned}
        \left[ n^+_{l,m}, n^-_{l',m'} \right] = &\sum_{\substack{L = 0 \\ l+l'+L \text{ even}}} h_L(s,l,l') (-1)^{m+m'} \begin{pmatrix}
            l & l' & L \\ m & m' & -m-m'
        \end{pmatrix}  n^{z}_{L,m+m'} \\
        - &\sum_{\substack{L = 0 \\ l+l'+L \text{ odd}}} h_L(s,l,l') (-1)^{m+m'} \begin{pmatrix}
            l & l' & L \\ m & m' & -m-m'
        \end{pmatrix}  n^{0}_{L,m+m'}.
    \end{aligned}
\end{equation}
Following \cite{He2025scalar}, we consider the subspace spanned by states with a small number \(k\) of spin flips and bounded angular momentum,
\begin{equation}\label{eq:oscillatorspace}
\prod_{i=1}^k n^+_{l_i,m_i}\ket{\Downarrow},
\qquad l_i \leq l_\text{max} .
\end{equation}
On this subspace, the density mode $n^z_{0,0}$ is extensive, with eigenvalue $-N+2k$ on a state with $k$ spin flips. 
In contrast, for $l > 0$ the density modes satisfy the operator norm bound (see Appendix~\ref{sec:oscillatorappendix} for derivation)
\begin{align*}
    \| n^z_{l,m} \|_{P_k} \leq 2k \sqrt{2l+1},
\end{align*}
where the $P_k$ subscript denotes the restriction to the sector with $k$ up spins.
The same bound holds for $n^0_{l,m}$.

Although \eqref{eq:commutatornplusminus} contains many $n^z_{L,m}$ terms, on the subspace of $k$ spin flips with $l_i \leq l_\text{max}$ only those with $L \leq 2kl_\text{max}$ survive. The reason is that $n^z_{L,m}$ carries angular momentum $L$, so for $L > 2kl_\text{max}$ it necessarily takes any state inside the subspace \eqref{eq:oscillatorspace} to states with angular momentum greater than $kl_\text{max}$ outside the subspace. 
Therefore, the operator norm of the remaining contribution can be bounded as 
\begin{align*}
    &\left\| \sum_{L =1}^{2kl_\text{max}} h_L(s,l,l') (-1)^{m+m'} \begin{pmatrix}
        l & l' & L \\ m & m' & -m-m'
    \end{pmatrix}n^z_{L,m+m'} \right\|_{P_k} \\ &\leq \sqrt{(2l+1)(2l'+1)} \, k^2 l_\text{max} \, C \exp{c \frac{(kl_\text{max})^2}{s}},
\end{align*}
for some constants $C$, $c$ independent of $s$. 
The inequality is derived in Appendix~\ref{sec:oscillatorappendix}.

In particular, if $l,l' \leq l_\text{max}$ and $k l_\text{max}=o(\sqrt{s})$, then the right-hand side divided by $N$ vanishes as $s\to\infty$. Then the rescaled modes satisfy the oscillator approximation on this subspace,
\begin{equation}
\Bigg[\frac{n^+_{l,m}}{\sqrt N},\,\frac{n^-_{l',m'}}{\sqrt N}\Bigg]
\;\approx\;
\delta_{l,l'}\,\delta_{m+m',0}\,\frac{n^z_{0,0}}{N}
\;\approx\;
\delta_{l,l'}\,\delta_{m+m',0}.
\end{equation}
Sharpening the bounds on the $3j$ and $6j$ symbols used in the derivation in Appendix~\ref{sec:oscillatorappendix} would likely improve the dependence on $k$ and $l_\text{max}$, but we do not pursue this here.

As an application of the oscillator approximation, we can estimate the error made by
replacing the exact density mode algebra by independent bosonic oscillators when evaluating observables in a variational state. 
We choose the state to be the wave function ansatz in \cite[Eq. 54]{He2025scalar}, given by
    \begin{align*}
        \ket{\psi} = \exp{S} \ket{\varphi_0}, \qquad S =  \sum_{l,m} c_l (-1)^m \frac{ n^-_{l,m} n^-_{l,-m} - n^+_{l,m} n^+_{l,-m}}{N},  
    \end{align*}
    where 
    \begin{align*}
        c_l =  \frac{1}{4} \log{\frac{u_l - v_l}{u_l + v_l}}, \qquad K_l = \sqrt{2s+1} \begin{pmatrix}
            s & s & l \\ s & -s & 0
        \end{pmatrix}.
    \end{align*}
    This ansatz has most of its weight in the subspace with a small number of spin-flips and is known to be close to the true ground state of the fuzzy sphere Hamiltonian for the free scalar CFT (cf. \cite[Table I, Fig. 8]{He2025scalar}).
    
    We will use the oscillator approximation to compute the correlation function
    \begin{align}\label{eq:correlatornx}
        \bra{\psi} n^x(\text{NP}) n^x(\gamma) \ket{\psi} = \frac{1}{N^2} \sum_{l,m,l',m'} Y_{l,m}(\text{NP}) Y_{l',m'}(\gamma) \, \bra{\psi} n^x_{l,m} n^x_{l',m'} \ket{\psi},
    \end{align}
    where NP denotes the north pole and $\gamma$ is short-hand for the point $(\gamma,0)$ on the sphere.
    The state $\ket{\psi}$ is chosen so that it is annihilated by the Bogoliubov rotated modes
    \begin{align*}
        \overline{a}_{l,m} = u_l \frac{(-1)^m n^-_{l,-m}}{\sqrt{N}} - v_l \frac{n^+_{l,m}}{\sqrt{N}},
    \end{align*}
    when the $n^+_{l,m}/\sqrt{N}$ are treated as independent harmonic oscillators. 
    In this approximation, the correlation function of the density modes evaluates to
    \begin{align*}
        \bra{\psi} n^x_{l,m} n^x_{l',m'} \ket{\psi} &= N e^{-2c_l} e^{-2c_{l'}} \bra{\psi} \left(\overline a^\dagger_{l,m} + (-1)^m \overline a_{l,-m}\right) \left(\overline a^\dagger_{l', m'} + (-1)^{m'} \overline a_{l',-m'} \right) \ket{\psi} \\
        & \approx N e^{-2c_\ell} e^{-2c_{\ell'}}
        (-1)^m
        \delta_{\ell\ell'}\delta_{m+m',0}.
    \end{align*}
    The first line rewrites the $n^x_{l,m}$ in terms of the Bogoliubov modes, and the second line uses the oscillator approximation to evaluate the contraction. This step is only meant to work for low-momentum modes with $l,l' \ll \sqrt{s}$, so that the calculation stays within the subspace where the approximation applies.
    
    Plugging the result back into \eqref{eq:correlatornx} gives 
    \begin{align*}
        \bra{\psi} n^x(\text{NP}) n^x(\gamma) \ket{\psi} = \sum_{l,m} e^{-4c_l} \frac{1}{N} \overline{Y}_{l,m}(\text{NP}) Y_{l,m}(\gamma) = \sum_{l,m} \frac{1}{K_l(2l+1)} \frac{1}{N} \overline{Y}_{l,m}(\text{NP}) Y_{l,m}(\gamma)
    \end{align*}
    The $l$-dependent coefficients in the sum should be compared with the CFT prediction and the coefficients measured numerically in the fuzzy sphere ground state: 
    \begin{align*}
        A_l^\text{CFT} = \frac{1}{2l+1}, \qquad A_l^\text{fuzzy ED} \approx \frac{K_l}{2l+1}, \qquad A_l^\text{osc} = \frac{1}{K_l(2l+1)}.
    \end{align*}
    The fuzzy sphere numerics converge to the CFT prediction multiplied with the UV regulator $K_l$ as $s \to \infty$, while the oscillator approximation gives the inverse factor. 
    These agree in the regime where the oscillator approximation is valid: for fixed $l \ll \sqrt{s}$, one has $K_l =1+\mathcal{O}(l^2/s)$. 

\section{Hopf algebra construction of \texorpdfstring{$so(3,2)$}{so(3,2)} representations}\label{sec: Hopf}

Establishing analytically how conformal symmetry emerges in the low-energy continuum limit of the fuzzy sphere Hamiltonian appears too difficult. Instead, we ask a much simpler question: does there exist an $so(3,2)$ representation whose generators are density modes $n^A_{l,m}$? We find that such a representation exists in the minimal nontrivial case of $s=\tfrac{1}{2}$, but does not persist for $s > \tfrac{1}{2}$. 
One can nevertheless extend the $s=\tfrac{1}{2}$ representation to larger systems via a Hopf algebra coproduct. However, this coproduct maps a spin-$s$ irrep of $so(3)$ to a tensor product of spin-$k$ and spin-$l$ irreps, so it splits a single fuzzy sphere into two fuzzy spheres. Therefore, it does not capture the thermodynamic limit relevant for our critical Hamiltonians, where one increases $s$ to obtain a single larger fuzzy sphere. 

\subsection{\texorpdfstring{$so(3,2)$}{so(3,2)} algebra representation of \texorpdfstring{$s=\tfrac{1}{2}$}{s=1/2} density modes}\label{sec:s12rep}

We identify the rotation generators of the $so(3,2)$ algebra as usual with
\begin{align*}
    J_z &= \frac{-2}{\sqrt{3}} n^0_{1,0}, \qquad J_\pm = \mp 2 \sqrt{\frac{2}{3}} n^0_{1,\pm1}.
\end{align*}
Then we propose the following identifications for the $so(2,1)$ generators (as defined in Section~\ref{subsec:conformalgenerators}):
\begin{align*}
    D_z = -3n^z_{0,0}, \quad D_+ = i \sqrt{3} n^+_{1,0}, \quad D_- = i \sqrt{3} n^-_{1,0}.
\end{align*}
Note that the $so(2,1)$ algebra acts on the spin degrees of freedom, corresponding to $\sigma^z$ and $i \sigma^\pm$ operators.
The remaining four ladder operators follow as 
\begin{align*}
    R_\pm = \pm [J_\pm, D_\pm] = \mp i \sqrt{6} n^\pm_{1,\pm1}, \quad S_\pm = \pm[J_\mp, D_\pm] = \mp i \sqrt{6} n^{\pm}_{1,\mp1}.
\end{align*}
It is straightforward to check that the defining $so(3,2)$ commutation relations \eqref{eq:so32commrel} are satisfied.

The $so(3,2)$ representation organizes the spectrum of the $D_z$ Hamiltonian in the 6d subspace with filling $N_\text{el}=2$ as follows: There is a 3d tower of states generated from the ground state $\ket{\uparrow \uparrow}$:
\begin{align*}
    \ket{\uparrow \uparrow}, E=-6 \quad \xrightarrow{D_-} \quad \ket{\uparrow \downarrow} - \ket{\downarrow \uparrow}, E=0 \quad \xrightarrow{D_-} \quad \ket{\downarrow \downarrow} E = 6.
\end{align*}
The other three states $\ket{\uparrow \downarrow} - \ket{\downarrow \uparrow}$, $\ket{e, \uparrow \downarrow}$, $\ket{\uparrow \downarrow, e}$, which form the $j=1$ rotation multiplet, are annihilated by $D_-$. 
Thus, in this tiny toy system, the 'conformal multiplet' structure contains two highest-weight states (the ground state with energy $-6$ and a $j=1$ states with energy $0$); the other states follow from acting with $D_-$ and $J_\pm$.

To obtain a unitary $so(5)$ representation instead of the non-unitary $so(3,2)$ one, it is enough to remove the factors of $i$ in $D_\pm$. With
\begin{align*}
    D_z = -n^z_{0,0}, \quad D_+^{so(5)} = \sqrt{3} n^+_{1,0}, \quad D_-^{so(5)} = \big(D_-^{so(5)}\big)^\dagger = \sqrt{3} n^-_{1,0},
\end{align*}
the generators $D_z$ and $D_\pm$ now close into an $so(3)$ subalgebra.
The $R^{so(5)}_\pm$ and $S^{so(5)}_\pm$ operators are then defined by the same commutation relations as before, but with $D_\pm$ replaced with $D_\pm^{so(5)}$.

In Appendix~\ref{sec:nogo}, we show that this $so(3,2)$/$so(5)$ representation built from the $n^A_{l,m}$ density modes at $s=\tfrac{1}{2}$ does not extend to larger $s$, even if we allow an arbitrary enlargement of the internal spin degrees of freedom. For example, it is easy to see that the commutator $[D_{+},D_-]$ produces an additional $n^z_{2,0}$ term when $s > \tfrac{1}{2}$.

\subsection{Larger \texorpdfstring{$so(3,2)$}{so(3,2)} representation from \texorpdfstring{$so(3)$}{so(3)} equivariant coproduct}\label{sec:so3coproduct}

To relate the fuzzy sphere algebras at different sizes $s$, we use the $so(3)$ equivariant coproduct of \cite{Balachandran:2003wv}. 
Let $\rho^J$ denote the spin-$J$ representation of $so(3)$ on the vector space $V^J \cong \mathbb{C}^{2J+1}$. For $M\in\text{End}(V^J)$ with matrix elements $M_{ij}$ in the weight basis $\{\ket{J,i}\}$, the coproduct is defined by
\begin{align*}
    \Delta_{KL}(M) = \Delta_{KL}\left(\sum_{ij} M_{ij} \ket{J,i} \bra{J,j}\right)= \sum_{k,l,k',l'} C^{Ji}_{K k, L l} C^{Jj}_{K k', L l'} M_{ij} \; \ket{K,k}\bra{K,k'} \otimes \ket{L,l} \bra{L,l'},
\end{align*}
where $C^{Ji}_{Kk,Ll}$ are Clebsch-Gordon coefficients. 
The coproduct is nonzero only when $\rho^J$ appears in the tensor product $\rho^K \otimes \rho^L$ (equivalently $|K-L| \leq J \leq K+L$). It is a homomorphism, i.e. $\Delta_{KL}(M_1 M_2) = \Delta_{KL}(M_1) \Delta_{KL}(M_2)$.

The coproduct is $so(3)$-equivariant with respect to the left action: for any generator $X\in so(3)$ and any $M\in \text{Mat}(2J{+}1)$,
\begin{align*}
\Delta_{KL}\!\big(\rho^{J}(X)\,M\big)
=\Big(\rho^{K}(X)\otimes \mathbb{I}+\mathbb{I}\otimes \rho^{L}(X)\Big)\,\Delta_{KL}(M),
\end{align*}
and analogously for the right action. 
Equivariance implies that the coproduct of an $so(3)$ generator itself is 
\begin{align*}
\Delta_{K,L}\!\left(\rho^J(X)\right)
&=
P_J \Big(
\rho^{K}(X)\otimes \mathbb{I}
+
\mathbb{I}\otimes \rho^{L}(X)
\Big)P_J, \quad \text{where } P_J=\Delta_{K,L}(\mathbb{I})
\end{align*}
is the projector onto the spin-$J$ irrep inside $K\otimes L$. The equivariant coproduct of the rotation generators is thus the usual Lie coproduct, followed by a projection onto the spin $J$ channel.
One can also define a counit and antipode, leading to a Hopf algebra \cite{Balachandran:2003wv} (strictly speaking, only when $J=K=L$).
 
To apply $\Delta_{KL}$ to the density modes $n^A_{l,m}$, we write them in their fermionic representation \eqref{eq:densitymodesTlm} and apply the coproduct to the orbital matrix. 
For our purposes, we choose $J=\tfrac{1}{2}$, $K=s$, and $L=s+\tfrac12$, so that $\Delta_{s,s+\frac12}$ maps an operator on the spin-$1/2$ fuzzy sphere to an operator on a spin-$(s+\tfrac12)$ fuzzy sphere coupled to a spin-$s$ sphere:
\begin{align*}
    \Delta_{s,s+1/2}(n^A_{l,m}) = \sum_{m_1,m_2=-s}^s \; \sum_{n_1,n_2=-s-1/2}^{s+1/2} c^\dagger_{m_1 n_1,\sigma} \, \big(\Delta_{s,s+1/2}(\hat{Y}_{l,m})\big)_{m_1 n_1,m_2 n_2} \, c_{m_2 n_2,\lambda} A_{\sigma,\lambda}. 
\end{align*}
While this coproduct gives us a $so(3,2)$ representation defined on a larger Hilbert space, it is different from the thermodynamic limit of the critical fuzzy sphere models, which takes $s \to \infty$ within a single $V^s$. Using the standard coproduct of Lie algebras leads to a similar mismatch, see Appendix~\ref{sec:liealgebracoproduct}. Nevertheless, it seems interesting that this coproduct induces a nontrivial $so(3,2)$ representation on a tensor product Hilbert space.

\section*{Acknowledgments}
We thank Yin-Chen He for his insightful comments on an early draft.  We thank the anonymous referees for their careful reading and helpful comments.
Z.W. is partially supported by ARO MURI contract W911NF-20-1-0082.  L.E. is supported by
the Walter Burke Institute for Theoretical Physics at
Caltech. This research was supported in part by grant NSF PHY-2309135 to the Kavli Institute for Theoretical Physics (KITP).

\vspace{1.8em}

\noindent{\bf \LARGE Appendices}

\appendix

\section{Derivation of the fermionic field anti-commutator}\label{sec:Kderivation}

Here we derive the anti-commutator of the fermionic fields \eqref{eq:fermionfieldLLL}. By definition of the fermionic fields \eqref{eq:fermionfieldLLL}, we have
\begin{align*}
    &\{\psi_\sigma(\theta, \phi), \psi^\dagger_\sigma(\theta, \phi')\} = \frac{1}{2s+1} \sum_{m=-s}^s Y^{(s)}_{s,m}(\theta, \phi) \left( Y^{(s)}_{s,m}(\theta', \phi') \right)^* \\
    &= \sum_{m=-s}^s \frac{(2s)!}{(s+m)!(s-m)!} e^{im(\phi - \phi')} \left( \cos\left( \frac{\theta}{2}\right) \cos\left( \frac{\theta'}{2}\right) \right)^{s+m} \left( \sin\left( \frac{\theta}{2}\right) \sin\left( \frac{\theta'}{2}\right) \right)^{s-m} \\ 
    &= e^{-is(\phi-\phi')} \sum_{k=0}^{2s} \frac{(2s)!}{k!(2s-k)!} \left( e^{i(\phi - \phi')} \cos\left( \frac{\theta}{2}\right) \cos\left( \frac{\theta'}{2}\right) \right)^{k} \left( \sin\left( \frac{\theta}{2}\right) \sin\left( \frac{\theta'}{2}\right) \right)^{2s-k} \\ 
    &=  e^{-is(\phi-\phi')} \sum_{k=0}^{2s} \binom{2s}{k} \left( e^{i(\phi - \phi')} \cos\left( \frac{\theta}{2}\right) \cos\left( \frac{\theta'}{2}\right) \right)^{k} \left( \sin\left( \frac{\theta}{2}\right) \sin\left( \frac{\theta'}{2}\right) \right)^{2s-k} \\
    &= e^{-is(\phi-\phi')} \left(e^{i(\phi-\phi')} \cos\left( \frac{\theta}{2}\right) \cos\left( \frac{\theta'}{2}\right) + \sin\left( \frac{\theta}{2}\right) \sin\left( \frac{\theta'}{2}\right) \right)^{2s}
\end{align*}
In the third line, we changed variables from $m$ to $k=s+m$, in the fourth line, we used the definition of $\binom{2s}{k}$, and in the fifth line, we used the identity $(x+y)^n=\sum_{k=0}^n \binom{n}{k} x^k y^{n-k}$. 
The magnitude of the term raised to the power of $2s$ is $\cos{(\gamma/2)}$, with $\gamma$ defined through
\begin{align*}
    \cos{\gamma} = \cos{\theta} \cos{\theta'} + \sin{\theta} \sin{\theta'} \cos{(\phi-\phi')}. 
\end{align*}
The phase $e^{i\chi(\theta,\phi,\theta',\phi')}$ of this term can be derived as 
\begin{align*}
    \arctan{\chi} = \frac{\sin{(\phi-\phi')} \cos{\left( \frac{\theta}{2} \right)} \cos{\left( \frac{\theta'}{2} \right)}}{\cos{(\phi-\phi')} \cos{\left( \frac{\theta}{2} \right)} \cos{\left( \frac{\theta'}{2} \right)} + \sin{\left( \frac{\theta}{2} \right)}  \sin{\left( \frac{\theta'}{2} \right)} },
\end{align*}
where the numerator is the imaginary part of the term and the denominator is the real part. 
So overall, we have 
\begin{align*}
    \{\psi_\sigma(\theta, \phi), \psi^\dagger_\sigma(\theta, \phi')\} = e^{2is \alpha(\theta, \phi, \theta', \phi')} \cos^{2s}{\left( \frac{\gamma}{2} \right)}, \quad \alpha(\theta, \phi, \theta', \phi') = -\frac{\phi-\phi'}{2} + \chi(\theta,\phi,\theta',\phi').
\end{align*}
The real phase $\alpha$ is anti-symmetric under exchanging $\theta \leftrightarrow \theta'$ and $\phi \leftrightarrow \phi'$.

\section{Derivation of the density mode commutator using fuzzy spherical harmonics}\label{sec:commutatorderivation}

To derive the density mode commutator \eqref{eq:modescommutator}, it is helpful to rewrite the density modes as in \eqref{eq:densitymodesTlm},
\begin{equation*}
n^A_{l,m} = \mathcal{N}_{s,l} \sum_{m_1, m_2=-s}^s \left(\hat{Y}_{l,m}\right)_{m_1,m_2} c^\dagger_{m_1,\sigma} A_{\sigma\lambda}c_{m_2,\lambda}, \quad \text{with } \mathcal{N}_{s,l} = (2s+1) \begin{pmatrix} s & s & l \\ s & -s & 0\end{pmatrix}.
\end{equation*}
The fuzzy spherical harmonics $\hat{Y}_{l,m}$ satisfy the algebra \cite{Chu:2001xi}
\begin{equation}\label{eq:Tlmproduct}
\begin{aligned}
    &\hat{Y}_{l,m} \hat{Y}_{l',m'} = \sum_{L,M} \,  A^{l,l',L}_{m,m',M} \hat{Y}_{L,M}, \\
    &\text{with } A^{l,l',L,s}_{m,m',M} = \sqrt{2s+1}(-1)^{2s+l+l'+M} \sqrt{(2l+1)(2l'+1)(2L+1)}
    \begin{pmatrix}
        l & l' & L \\ m & m' & -M
    \end{pmatrix}  \begin{Bmatrix}
        l & l' & L \\ s & s & s
    \end{Bmatrix},
\end{aligned}
\end{equation}
where the coefficients in the expansion match the coefficients in the product of two spherical harmonics $Y_{l,m}$. 
Using the identity 
\begin{align*}
    \left[c^\dagger_{m_1,\sigma} c_{m_2,\lambda}, c^\dagger_{m'_1,\sigma'} c_{m'_2,\lambda'},\right] = \delta_{\lambda \sigma'} \delta_{m_2,m'_1} c^\dagger_{m_1,\sigma} c_{m'_2,\lambda'} - \delta_{\sigma \lambda'} \delta_{m_1,m'_2} c^\dagger_{m'_1,\sigma'} c_{m_2,\lambda},
\end{align*}
the commutator of the $n^A_{l,m}$ density modes can be expanded as 
\begin{equation*}
    \begin{aligned}
        \left[n^A_{l,m},n^B_{l',m'}\right] = \mcN_{s,l} \mcN_{s,l'} \Bigg( &\sum_{m_1, m_2, m'_2} &(\hat{Y}_{l,m})_{m_1,m_2}  (\hat{Y}_{l',m'})_{m_2,m'_2} c^\dagger_{m_1,\sigma} (A \cdot B)_{\sigma \lambda'} c_{m'_2,\lambda'} \\
        - &\sum_{m_1, m_2, m'_1} & (\hat{Y}_{l',m'})_{m'_1,m_1} (\hat{Y}_{l,m})_{m_1,m_2}  c^\dagger_{m'_1,\sigma'} (B \cdot A)_{\sigma' \lambda} c_{m_2,\lambda} \Bigg).
    \end{aligned}
\end{equation*}
Splitting $A\cdot B$ and $B \cdot A$ into commutator and anti-commutator and relabeling dummy indices gives 
\begin{equation}\label{eq:nlm_comm_Tlm}
    \begin{aligned}
        \left[n^A_{l,m},n^B_{l',m'}\right] =  \mcN_{s,l} \mcN_{s,l'} \Bigg( \sum_{m_1, m_2} &\left[\hat{Y}_{l,m}, \hat{Y}_{l',m'} \right]_{m_1,m_2} c^\dagger_{m_1,\sigma} \{A,  B\}_{\sigma \lambda} c_{m_2,\lambda} \\
        &+ \left\{ \hat{Y}_{l,m}, \hat{Y}_{l',m'} \right\}_{m_1,m_2} c^\dagger_{m_1,\sigma} [A,  B]_{\sigma \lambda} c_{m_2,\lambda} \Bigg)
    \end{aligned}
\end{equation}
Now note that in the product of the $\hat{Y}_{l,m}$ matrices in \eqref{eq:Tlmproduct}, the 6j symbol is invariant under exchanging $l \to l'$ while the 3j symbol picks up a $(-1)^{l+l'+L}$ sign. Therefore, 
\begin{equation*}
    \begin{aligned}
        \left[n^A_{l,m},n^B_{l',m'}\right] =  \mcN_{s,l} \mcN_{s,l'} \sum_{m_1, m_2} \sum_{L,M} \Bigg(& \left(1 - (-1)^{l+l'+L}\right) A^{l,l',L,s}_{m,m',M} (\hat{Y}_{L,M})_{m_1,m_2} c^\dagger_{m_1,\sigma} \{A,  B\}_{\sigma \lambda} c_{m_2,\lambda} \\
        &+ \left(1 + (-1)^{l+l'+L}\right) A^{l,l',L,s}_{m,m',M} (\hat{Y}_{L,M})_{m_1,m_2} c^\dagger_{m_1,\sigma} [A,  B]_{\sigma \lambda} c_{m_2,\lambda} \Bigg)
    \end{aligned}
\end{equation*}
The $\{A,  B\}$ anticommutator only has support when $l+l'+L$ is odd, while the $[A,  B]$ commutator only has support when $l+l'+L$ is even. After inserting the expressions for $\mcN_{s,l}$ and $A^{l,l',L,s}_{m,m',M}$, this reduces to the commutator \eqref{eq:modescommutator}.

\section{Derivation of oscillator approximation}\label{sec:oscillatorappendix}

Here we derive the operator norm bounds for the oscillator approximation in Section~\ref{subsec:oscillator}. Because $n^z_{l,m}$ and $n^0_{l,m}$ conserve the number of spin-flips $k$, it suffices to work in a fixed-$k$ sector. In the reference state ($k=0$), we have $\bra{\Downarrow} n^z_{l,m} \ket{\Downarrow} = -N \delta_{l,0}$ (similarly $\bra{\Downarrow} n^0_{l,m} \ket{\Downarrow} = N \delta_{l,0}$.  

In the $k=1$ subspace spanned by states $\ket{a;b} := c^\dagger_{a,\uparrow} c_{b,\downarrow} \ket{\Downarrow}$, the matrix elements of $n^z_{l,m}$ with $l > 0$ are 
\begin{align*}
    \bra{a';b'} n^z_{l,m} \ket{a;b} &= \mcN_{s,l} \sum_{m_1,m_2} (\hat{Y}_{l,m})_{m_1,m_2} \bra{a';b'} c^\dagger_{\uparrow,m_1} c_{\uparrow, m_2} - c^\dagger_{\downarrow,m_1} c_{\downarrow, m_2} \ket{a; b} \\
    &= \mcN_{s,l} \left( (\hat{Y}_{l,m})_{a',a} \delta_{b,b'} + (\hat{Y}_{l,m})_{b,b'} \delta_{a,a'}  \right) 
\end{align*}
Then using the triangle inequality for the two terms, we can bound the operator norm of $n^z_{l,m}$ with $l > 0$ in this subspace by
\begin{align*}
    \| n^z_{l,m} \| \leq 2 \mcN_{s,l} \| \hat{Y}_{l,m} \|.
\end{align*}
To bound the operator norm of $\hat{Y}_{l,m}$, we use that the norm is equal to the square root of the largest eigenvalue of $\hat{Y}^\dagger_{l,m} \hat{Y}_{l,m}$. This is a diagonal matrix,
\begin{align*}
    (\hat{Y}^\dagger_{l,m} \hat{Y}_{l,m})_{bb'} = \sum_a (\hat{Y}_{l,m})^*_{ab} (\hat{Y}_{l,m})_{ab'} = \delta_{b,b'} \sum_a |(\hat{Y}_{l,m})_{ab}|^2 = \delta_{b,b'} |(\hat{Y}_{l,m})_{b+m,b}|^2,  
\end{align*}
because $(\hat{Y}_{l,m})_{ab}$ is nonzero only when $b=m-a$ and $(\hat{Y}_{l,m})_{ab'}$ is nonzero only when $b'=m-a$. Therefore, the operator norm of $\hat{Y}_{l,m}$ is equal to 
\begin{align*}
    \| \hat{Y}_{l,m} \| = \max_{b}|(\hat{Y}_{l,m})_{b+m,b}| = \sqrt{2l+1} \max_b \left| \begin{pmatrix}
        s & s & l \\ b+m & -b & -m
    \end{pmatrix} \right| \leq \sqrt{\frac{2l+1}{2s+1}}.
\end{align*}
The last inequality follows from the condition 
\[
(2j_3+1) \sum_{m_1,m_2} \begin{pmatrix}
    j_1 & j_2 & j_3 \\ m_1 & m_2 & m_3 \end{pmatrix}^2 = \begin{cases}
        1 & |j_1 - j_2 | \leq j_3 \leq j_1 + j_2 \\ 
        0 & \text{else } 
    \end{cases},
\] 
which implies 
\begin{align*}\label{eq:3jsymbolbound}
\begin{pmatrix}
    j_1 & j_2 & j_3 \\ m_1 & m_2 & m_3 \end{pmatrix}^2 \leq \frac{1}{2\max{\{j_1,j_2,j_3\}}+1}.
\end{align*}
Together with the fact that $\mcN_{s,l} = (2s+1) \begin{pmatrix}
    s & s & l \\ s & - s & 0
\end{pmatrix} \leq \sqrt{2s+1}$, this leads to
\begin{align*}
    \| n^z_{l,m} \| \leq 2 \sqrt{2l+1}.
\end{align*}

In contrast, for $n^z_{0,0}$ we have 
\[
    n^z_{0,0} \ket{a;b} = (N-2) \ket{a;b} \quad \Rightarrow \quad \| n^z_{0,0} \| = N-2,
\]
therefore $n^z_{0,0}$ is extensive in $N$ while $\| n^z_{l,m} \| = \mcO(\sqrt{l})$ for $l > 0$ (and analogously for $n^0_{l,m}$). 

In the sector with $k$ spin flips spanned by states $\ket{A;B} := c^\dagger_{a_1} \dots c^\dagger_{a_k} c_{b_1} \dots c_{b_k} \ket{\Downarrow}$ with $A = \{a_1, \dots, a_k\}$ and $B = \{b_1, \dots, b_k\}$, the matrix elements of $n^z_{l,m}$ are
\begin{align*}
    \bra{A';B'} n^z_{l,m} \ket{A;B} = \mcN_{s,l} \sum_{i=1}^k \left( (\hat{Y}_{l,m})_{a'_i,a_i} \delta_{A \setminus a_i, A' \setminus a'_i} \delta_{B,B'} + (\hat{Y}_{l,m})_{b_i,b'_i} \delta_{B \setminus b_i, B' \setminus b'_i} \delta_{A,A'} \right)
\end{align*}
Using again the triangular inequality for the $2k$ terms, the operator norm is then bounded as 
\begin{align*}
    \| n^z_{l,m} \| \leq 2k \mcN_{s,l} \| \hat{Y}_{l,m} \| \leq 2k \sqrt{2l+1}.
\end{align*}
Again, the same holds for $n^0_{l,m}$.
In contrast, the operator norm of $n^z_{0,0}$ is again extensive, concretely 
\[
n^z_{0,0} \ket{A;B} = (-N+2k) \ket{A;B} \quad \Rightarrow \| n^z_{0,0} \| = N-2k. 
\]

Next we estimate the operator norm of the total contribution after further restricting to the subspace $\prod_{i=1}^k n^+_{l_i,m_i} \ket{\Downarrow}$ with $l_i \leq l_\text{max}$. As explained in Section~\ref{subsec:oscillator}, this restriction truncates the sum over $L$ at $L=2kl_\text{max}$. Using the Cauchy-Schwartz inequality,
\begin{align*}
    &\left\| \sum_{L =1}^{2kl_\text{max}} h_L(s,l,l') (-1)^{m+m'} \begin{pmatrix}
        l & l' & L \\ m & m' & -m-m'
    \end{pmatrix}n^z_{L,m+m'} \right\|_{P_k} \\ &\leq \left(\sum_{L =1}^{2kl_\text{max}} (h_L(s,l,l'))^2 \begin{pmatrix}
        l & l' & L \\ m & m' & -m-m'
    \end{pmatrix}^2 \right)^{1/2} \left( \sum_{L=1}^{2kl_\text{max}} \| n^z_{L,m+m'} \|^2_{P_k} \right)^{1/2}
\end{align*}
To bound the first factor, we use $\begin{pmatrix}
    j_1 & j_2 & j_3 \\ m_1 & m_2 & m_3
\end{pmatrix} \leq \frac{1}{\sqrt{2\max{\{j_1,j_2,j_3 \}}+1}}$ on three different 3j symbols, 
\begin{align*}
\sum_{L =1}^{2kl_\text{max}} (h_L(s,l,l'))^2 \begin{pmatrix}
        l & l' & L \\ m & m' & -m-m'
    \end{pmatrix}^2 
    \leq (2l+1)(2l'+1) \sum_{L=1}^{2kl_\text{max}}\frac{\begin{Bmatrix} l & l' & L \\ s & s & s \end{Bmatrix}^2}{ \begin{pmatrix} s & s & L \\ s & -s & 0 \end{pmatrix}^2 }.
\end{align*}
Orthogonality of the 6j symbols then gives
\begin{align*}
    \sum_{L=1}^{2kl_\text{max}}\frac{\begin{Bmatrix} l & l' & L \\ s & s & s \end{Bmatrix}^2}{ \begin{pmatrix} s & s & L \\ s & -s & 0 \end{pmatrix}^2 } \leq \frac{1}{2s+1} \max_{L=1,\dots,2kl_\text{max}} \begin{pmatrix}
        s & s & L \\ s & -s & 0
    \end{pmatrix}^{-2}
\end{align*}
The inverse squared 3j symbol is known to be
\begin{align*}
    \begin{pmatrix}
        s & s & L \\ s & -s & 0
    \end{pmatrix}^{-2} = \frac{(2s-L)!(2s+L+1)!}{((2s)!)^2}.
\end{align*}
For $L_\text{max} = 2kl_\text{max} \leq s/2$, this can be bounded using Stirling's approximation as
\begin{align*}
\max_{L=1,\dots,2kl_\text{max}} 
    \begin{pmatrix}
        s & s & L \\ s & -s & 0
    \end{pmatrix}^{-2} = \frac{(2s-L_\text{max})!(2s+L_\text{max}+1)!}{((2s)!)^2} \leq s C \exp{\frac{cL_\text{max}^2}{s}},
\end{align*}
where $c$ and $C$ are some constants. If $L_\text{max} \ll \sqrt{s}$, this exponential is $\mcO(1)$. Altogether, this gives the bound
\begin{align*}
    &\left\| \sum_{L =1}^{2kl_\text{max}} h_L(s,l,l') (-1)^{m+m'} \begin{pmatrix}
        l & l' & L \\ m & m' & -m-m'
    \end{pmatrix}n^z_{L,m+m'} \right\|_{P_k} \\ &\leq \sqrt{(2l+1)(2l'+1)} \; C \exp{\frac{c \, k^2 l_\text{max}^2}{s}} \; k^2 l_\text{max}
\end{align*}

\section{Central extensions}\label{sec:centralextension}

The planar GMP algebra admits a central extension, giving rise to the W-infinity algebra \cite{Cappelli:2021kxd}. 
The Lie algebra of our density modes $\{n^A_{l,m}\}_{l \geq 1}$ at fixed $s$ is finite-dimensional; assuming it is semisimple as well, Whitehead’s lemma implies that any putative central extension is cohomologically trivial. Nevertheless, it is still natural to search for central terms at finite $s$ and ask whether they can survive in a suitable large-$s$ limit, potentially matching the central extension of area-preserving diffeomorphisms of $S^2$ discussed in \cite{Bas}.

Ignoring the spin dof (i.e. setting $A=\text{id}$ in all $n^A_{l,m}$), a central extension should take the form 
\begin{equation}\label{eq:centralextenxion}
    \begin{aligned}
        \left[ n_{l,m}, n_{l',m'} \right] = 
        - 2\hspace{-3mm} \sum_{\substack{L = 0 \\ l+l'+L \text{ odd}}} \hspace{-2mm} h_L(s,l,l') (-1)^{m+m'} \begin{pmatrix}
            l & l' & L \\ m & m' & -m-m'
        \end{pmatrix}  n_{L,m+m'} + \delta_{m+m',0} \; g(s,l,l',m) \mathbb{I}.
    \end{aligned}
\end{equation}
The $\delta_{m+m',0}$ delta function is necessary because of the grading of the algebra: $n_{1,0}$ acts as a grading operator, $[n_{1,0}, n_{l,m}] \propto m \, n_{l,m}$ (similar to $L_0$ in Virasoro). Consistency of the commutator with this grading requires the central extension term to vanish unless $m+m'=0$.

The function $g$ must satisfy $g(s,l,l',m) = -g(s,l',l,-m)$ so that $[n_{l,m}, n_{l',m'}] = -[n_{l',m'}, n_{l,m}]$ is not violated. Further constraints come from the Jacobi identity:
\begin{align*}
    \sum_{\substack{L = 0 \\ l_1+l_2+L \text{ odd}}} h_L(s,l_1,l_2) (-1)^{m_1+m_2} \begin{pmatrix}
        l_1 & l_2 & L \\ m_1 & m_2 & m_3
    \end{pmatrix} g(s,l_3,L,m_3) + \text{ cyclic permutations } = 0,
\end{align*}
assuming $m_1+m_2+m_3=0$. One possible solution that satisfies the Jacobi identity is 
\begin{align*}
    g(s,l,l',m) = \delta_{l,l'} (-1)^m \,m\begin{pmatrix}
        s & s & l \\ s & -s & 0
    \end{pmatrix}^2. 
\end{align*}
We leave it to future work to determine whether this solution is connected to the central extension of the planar GMP algebra in the $s \to \infty$ limit with $l \sim \sqrt{s}$, or related to other central extensions for 3d CFTs \cite{Bas}.

\section{No \texorpdfstring{$so(5)$}{so(5)} representation extending \texorpdfstring{$so(3)$}{so(3)} representation generators for \texorpdfstring{$s > \tfrac{1}{2}$}{s > 1/2}}\label{sec:nogo}

In this appendix we consider the general class of fermion-bilinear operators $n^A(X)$ and ask whether one can extend the orbital rotation generators $J_a = n^\mathbb{I}(L_a)$ to a larger $so(5)$ (or $so(3,2)$) Lie algebra. We show that for $s > \tfrac{1}{2}$ this is not possible, independent of the internal spin space dimension.

A fermion bilinear built from \(2s+1\) orbital modes and \(d\) internal spin/flavor degrees of freedom can be written as
\begin{align*}
    n^A(X)
    := \sum_{m_1,m_2=-s}^s \sum_{\sigma,\lambda=1}^d
    X_{m_1 m_2}\,
    c^\dagger_{m_1,\sigma}\, A_{\sigma\lambda}\, c_{m_2,\lambda},
\end{align*}
where \(X\) is a \((2s+1)\times(2s+1)\) matrix acting on the orbital indices \(m\), and \(A\) is a \(d\times d\) matrix acting on the internal indices \(\sigma\).
The density modes \(n^A_{l,m}\) in \eqref{eq:densitymodesTlm} are the special case \(X=\hat{Y}_{l,m}\) (the fuzzy spherical harmonics) with \(d=2\) corresponding to spin \(\uparrow,\downarrow\).
Using the canonical anticommutation relations for the fermions, the commutator of two such bilinears closes:
\begin{equation*}
\begin{aligned}
    \bigl[n^A(X),\, n^B(Y)\bigr]
    &= n^{AB}(XY) - n^{BA}(YX) = \frac12\, n^{\{A,B\}}\!\bigl([X,Y]\bigr)
      + \frac12\, n^{[A,B]}\!\bigl(\{X,Y\}\bigr).
\end{aligned}
\end{equation*}
The corresponding special case for \(n^A_{l,m}\) is \eqref{eq:nlm_comm_Tlm}.

The $so(3)$ rotation generators are
\begin{align*}
    J_z = n^{\mathbb{I}}(L_z), \qquad J_\pm = n^{\mathbb{I}}(L_\pm),
\end{align*}
where \(\mathbb{I}\) denotes the \(d\times d\) identity and \(L_z,L_\pm\) are the generators of the spin-$s$ irrep of $so(3)$.
We want to find an additional fermion-bilinear generator \(D_z=n^{A_z}(X_z)\) that commutes with all rotations.
Since \([L_a,D_z]=0\) implies \([L_a,X_z]=0\) for \(a=z,\pm\), Schur's lemma for the irreducible spin-$s$ orbital representation gives $X_z \propto \mathbb{I}_{2s+1}$, so we may take 
\begin{align*}
    D_z = n^{A_z}(\mathbb{I}) = \sum_{m_1=-s}^s c^\dagger_{m_1,\sigma} (A_z)_{\sigma\lambda} c_{m_1,\lambda}.
\end{align*}
In other words, rotational invariance restricts $D_z$ to an onsite term that acts independently on each orbital $m$.

Next, impose \([J_z,D_\pm]=0\).  Writing \(D_\pm=n^{A_\pm}(X_\pm)\), the commutator formula reduces to
\begin{align*}
    [J_z,D_\pm] = n^{A_\pm}([L_z,X_\pm]),
\end{align*}
so \([J_z,D_\pm]=0\) forces \([L_z,X_\pm]=0\), hence \(X_\pm\) are diagonal in the \(L_z\) eigenbasis \(\{\ket{m}\}\).
With \(X_\pm\) diagonal we also have \([X_+,X_-]=0\), and therefore
\begin{align*}
    [D_+,D_-]
    = \frac12\, n^{[A_+,A_-]}(\{X_+,X_-\}).
\end{align*}
Requiring \([D_+,D_-]\) to be proportional to \(D_z=n^{A_z}(\mathbb{I})\) implies
\begin{align*}
    [A_+,A_-] \propto A_z,
    \qquad
    \{X_+,X_-\} \propto \mathbb{I}_{2s+1}.
\end{align*}

If the Fock-space inner product is to be invariant for the $so(5)$ representation, we can choose conventions such that $D_- = (D_+)^\dagger$.
Since \((n^A(X))^\dagger = n^{A^\dagger}(X^\dagger)\), this gives \(A_- = A_+^\dagger\) and \(X_- = X_+^\dagger\).
Writing \(X_+=\mathrm{diag}(k_{-s},\dots,k_s)\), the condition \(\{X_+,X_-\}\propto \mathbb{I}\) becomes
\begin{align*}
    \{X_+,X_+^\dagger\} = 2\,\mathrm{diag}(|k_m|^2) \propto \mathbb{I}
    \quad\Longrightarrow\quad
    |k_m|=\text{const}.
\end{align*}
With the normalization used above, this is \(|k_m|=1/\sqrt2\), i.e. \(k_m=e^{i\theta_m}/\sqrt2\).

A further relation required by the $so(5)$ algebra in this ladder basis is
\begin{align*}
    [J_-,[J_-,D_+]]=0.
\end{align*}
Using the closure of bilinears and that \(J_-=n^{\mathbb{I}}(L_-)\), this reduces to 
\begin{align*}
    [J_-,[J_-,D_+]]
    = n^{A_+}\!\bigl([L_-,[L_-,X_+]]\bigr),
\end{align*}
so we must have \([L_-,[L_-,X_+]]=0\).
In the \(\ket{m}\) basis where \(X_+\) is diagonal, \(L_-\ket{m}=c_m\ket{m-1}\) with
\(c_m=\sqrt{(s+m)(s-m+1)}\). A direct computation gives
\begin{align*}
    [L_-,[L_-,X_+]]\ket{m}
    = c_m c_{m-1}\,\bigl(k_m - 2k_{m-1} + k_{m-2}\bigr)\,\ket{m-2}.
\end{align*}
For \(s>\tfrac12\), the coefficients \(c_m c_{m-1}\) are nonzero for \(m=-s+2,\dots,s\), hence
\begin{align*}
    k_m - 2k_{m-1} + k_{m-2} = 0
    \qquad (m=-s+2,\dots,s).
\end{align*}
Thus \(k_m\) has vanishing second finite difference and must be affine:
\begin{align*}
    k_m = a m + b.
\end{align*}
Combining this with \(|k_m|=1/\sqrt2\) for all \(m\) forces \(a=0\) whenever there are at least three distinct \(m\)-values, i.e. for \(s\ge 1\).
Hence \(k_m\) must be constant, so \(X_+\propto \mathbb{I}\) and therefore \(D_+\) commutes with all \(J_a\), contradicting the requirement that \(D_\pm\) provide a nontrivial extension of the rotation algebra.
The case \(s=\tfrac12\) evades this obstruction because \((L_-)^2=0\), so \([L_-,[L_-,X_+]]=0\) holds identically.

For \(s>\tfrac12\), there is no choice of diagonal \(X_+\) with \(|k_m|=\mathrm{const}\) that satisfies the mixed relation \([J_-,[J_-,D_+]]=0\) while keeping \(D_+\) nontrivial.
This obstruction is independent of the internal dimension \(d\).
In particular, there is no $so(5)$ representation on these fermion bilinears (and most likely also not $so(3,2)$ one) that extends the spin-$s$ orbital $so(3)$ generators and is unitary with respect to the Fock inner product for \(s>\tfrac12\).

\section{Larger \texorpdfstring{$so(3,2)$}{so(3,2)} representations from the Lie algebra coproduct}\label{sec:liealgebracoproduct}

As a complementary construction, one may enlarge the minimal $s=\tfrac12$ representation by using the standard coproduct on the universal enveloping algebra $U(so(3,2))$.  For any generator $x\in so(3,2)$ we define
\begin{align*}
    \Delta(x) := x\otimes \mathbb{I}+\mathbb{I}\otimes x ,
\end{align*}
and obtain an $n$-fold coproduct $\Delta^{(n)}$ by iteration. 

While this produces an $so(3,2)$ representation on a larger Hilbert space, it does not match the notion of ``increasing the fuzzy sphere size'' relevant for the critical models.  In particular, the induced $so(3)$ rotations act on the tensor product Hilbert space $\mathcal H_{s=\frac12}^{\otimes n}$ and therefore furnish the reducible representation $(V^{\frac12})^{\otimes n}$ rather than a single spin-$s$ irrep $V^s$ with $s=\tfrac{n}{2}$. Consequently, the rotation generators obtained from $\Delta^{(n)}$ are not proportional to the $l=1$ density modes $n^{0,s}_{1m}$ of a spin-$s$ fuzzy sphere. 

One may also deform the coproduct by a Drinfeld twist.  For an invertible $F\in H\otimes H$ with $H=U(so(3,2))$, the twisted coproduct is
\begin{align*}
    \Delta^F(x):=F\,\Delta(x)\,F^{-1},
\end{align*}
which still satisfies $\,[\Delta^F(x),\Delta^F(y)]=\Delta^F([x,y])\,$.  However, since $\Delta^F(x)$ is conjugate to $\Delta(x)$, twisting does not change the equivalence class of the resulting tensor-product representation. In particular, it cannot convert the coproduct rotations on $(V^{\frac12})^{\otimes n}$ into the single-irrep rotations appropriate to a spin-$s$ fuzzy sphere, and therefore does not resolve the structural mismatch with the thermodynamic limit of the critical fuzzy sphere models.

\small
\bibliographystyle{apsrev}
\bibliography{references}

\end{document}